\documentclass[twocolumn,epjc3,envcountsect]{svjour3}

\RequirePackage[T1]{fontenc}
\usepackage[american]{babel}
\usepackage[normalem]{ulem}

\usepackage{comment}

\RequirePackage[T1]{fontenc}

\smartqed

\RequirePackage{graphicx}
\RequirePackage{mathptmx}
\RequirePackage{flushend}
\RequirePackage[numbers,sort&compress]{natbib}
\RequirePackage[colorlinks,citecolor=blue,urlcolor=blue,linkcolor=blue]{hyperref}

\usepackage{amssymb}
\usepackage{color,colortbl}
\usepackage{epstopdf}
\usepackage{graphicx}
\usepackage{tabularx}
\usepackage{siunitx} 
\usepackage{booktabs}
\usepackage{amsmath}
\usepackage{dsfont}
\usepackage{xstring}
\usepackage{subcaption}

\usepackage{microtype} 

\usepackage{arydshln}

\usepackage[capitalize]{cleveref}
\Crefname{section}{Section}{Sections}
\Crefname{figure}{Figure}{Figures}
\Crefname{table}{Table}{Tables}
\Crefname{appendix}{Appendix}{Appendices}
\Crefname{equation}{Eq.}{Eqs.}
\newcommand{\Onlinecite}[1]{%
    \IfSubStr{#1}{,}{Refs}{Ref}.~\cite{#1}%
}

\definecolor{Gray}{gray}{0.9}

\journalname{Eur. Phys. J. C}

\begin{document}

\title{Hints for string breaking in QCD}

\author{
P. Cea\thanksref{e1,addr1}
\and
V. Chelnokov\thanksref{e2,addr2,addr3}
\and
L. Cosmai\thanksref{e3,addr1}
\and
A. Papa\thanksref{e4,addr4,addr5}
}

\institute{
INFN - Sezione di Bari, I-70126 Bari, Italy\label{addr1}
\and
Institut f\"ur Theoretische Physik, Goethe Universit\"at, 60438 Frankfurt am Main, Germany\label{addr2}
\and
Bogolyubov Institute for Theoretical Physics, National
Academy of Sciences of Ukraine, 03143 Kyiv, Ukraine\label{addr3}
\and
Dipartimento di Fisica, Universit\`a della Calabria, I-87036 Arcavacata di Rende, Cosenza, Italy\label{addr4}
\and
INFN - Gruppo collegato di Cosenza, I-87036 Arcavacata di Rende, Cosenza, Italy\label{addr5}
}

\thankstext{e1}{e-mail: paolo.cea@ba.infn.it}
\thankstext{e2}{e-mail: volodymyr.chelnokov@gmail.com}
\thankstext{e3}{e-mail: leonardo.cosmai@ba.infn.it}
\thankstext{e4}{e-mail: alessandro.papa@fis.unical.it}

\date{Received: date / Accepted: date}

\maketitle

\begin{abstract}
We present results for the chromo-electric field generated by a static quark–antiquark pair at nearly zero temperature in lattice QCD with 2+1 dynamical staggered fermions at physical quark masses. We investigate the evolution of the flux-tube structure as the distance between the static color charges increases. 
We find hints that string breaking occurs at a distance in the range  $0.963 \;  \text{fm} \;  \lesssim  \; d^* \lesssim \; 1.156 \; \text{fm}$. 
\end{abstract}

\section{Introduction}
\label{S1}
\begin{figure*}[htb]
\begin{center}
\includegraphics[width=0.47\linewidth,clip]{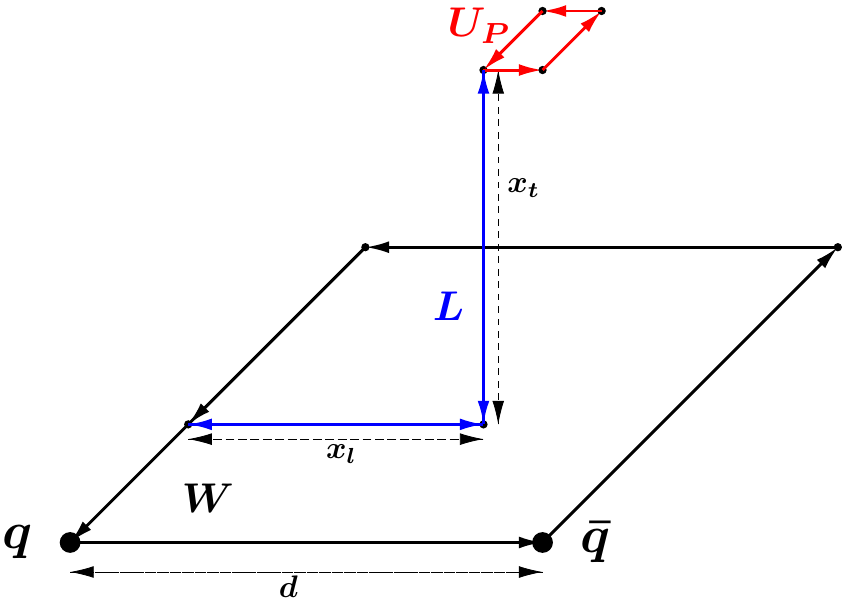}
\includegraphics[width=0.47\linewidth,clip]{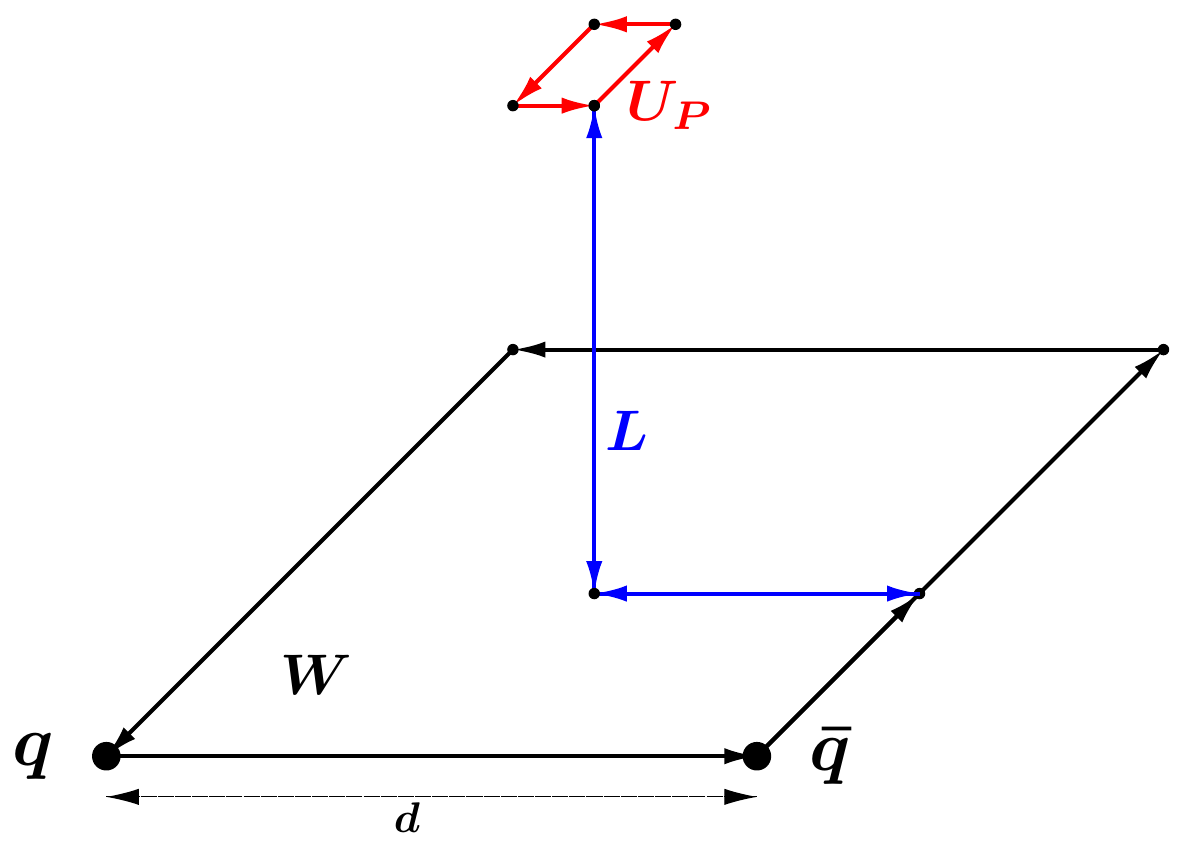}
\end{center}
\caption{The flux-tube operator with the Schwinger line attached to the quark time line (left) or to the antiquark time line (right).
}            
\label{fluxtubeoperator}
\end{figure*}

Quantum chromodynamics (QCD) is the theory of strong interactions
describing the dynamics of quarks and gluons (see Ref.~\cite{Gross:2022hyw} and references therein).
It is a well-established fact that the phenomenon of confinement of quarks and gluons inside hadrons is encoded into the QCD Lagrangian, mainly due to
 a wealth of numerical analyses of QCD on  a space-time lattice. Moreover, there are no doubts that   the chromo-electric field between two static quarks distributes 
 in tube-like  structures~\cite{Fukugita:1983du,Kiskis:1984ru,Flower:1985gs,Wosiek:1987kx,DiGiacomo:1989yp,DiGiacomo:1990hc,Singh:1993jj,Matsubara:1993nq,Cea:1992vx,Cea:1993pi,Cea:1994ed,Cea:1994aj,Cea:1995zt,Bali:1994de,Haymaker:2005py,D'Alessandro:2006ug,Cardaci:2010tb,Cea:2012qw,Cea:2013oba,Cea:2014uja,Cea:2014hma,Cardoso:2013lla,Caselle:2014eka,Bicudo:2017uyy,Bicudo:2018jbb}.
From these tube-like structures a linear potential between static color charges naturally arises, thus furnishing
the numerical evidence for quark confinement. \\
To reach a microscopic understanding of color confinement, in recent years we have undertaken numerical studies based on Monte Carlo simulations of the SU(3) Yang--Mills theory and of full QCD with (2+1) Highly Improved Staggered Quarks (HISQ)  discretized on a Euclidean space--time lattice. 
Within this framework, we have systematically analyzed the local structure of the color fields generated by two static color sources, a quark and an antiquark~\cite{Baker:2018mhw,Baker:2019gsi,Baker:2022cwb,Baker:2023dnn,Baker:2024peg,Baker:2024rjq,Baker:2025mec,Baker:2025bja}.
To this end, we extracted the spatial distribution of the color fields generated by a static quark--antiquark pair from lattice measurements of the connected correlation function \(\rho^{\text{conn}}_{W,\mu\nu}\). This correlator involves a plaquette \(U_P = U_{\mu\nu}(x)\) in the \(\mu\nu\) plane and a rectangular Wilson loop \(W\) connected by a Schwinger line $L$: 
\begin{equation}
    \rho^{\text{conn}}_{W,\mu\nu} = 
    \frac{\langle \mathrm{tr}(W L U_P L^\dagger)\rangle}{\langle \mathrm{tr}(W)\rangle}
    - \frac{1}{N}\,\frac{\langle \mathrm{tr}(U_P)\,\mathrm{tr}(W)\rangle}{\langle \mathrm{tr}(W)\rangle}\;,
    \label{connected1}
\end{equation}
where \(N=3\) is the number of QCD colors.  The Schwinger line can be attached to the quark time line or to the antiquark time line (see Fig.~\ref{fluxtubeoperator} and
 compare it with  Fig.~1 in Ref.~\cite{Baker:2025bja}).
The correlator $\rho^{\text{conn}}_{W,\mu\nu}$ provides a gauge-invariant lattice definition of a field-strength tensor $ F_{\mu\nu}$ induced by the static sources,
\begin{equation}
    \rho^{\text{conn}}_{W,\mu\nu} \;
    \;\equiv\; a^2 g\,F_{\mu\nu} \; ,
    \label{connected2}
\end{equation}
where the tensor $F_{\mu\nu}$  carries one unit of octet charge and depends linearly on the color components of the fields.
According to the orientation of the plaquette $U_P$ in Eq.~(\ref{connected1}), different field components are accessed.  \\
Our  previous numerical studies  can be summarized by the following key results. Firstly, 
 the chromo-magnetic field $\vec{B}$ induced by the static sources is statistically consistent with zero within our present accuracy.  
 In addition, the chromo-electric~\footnote{In the following we  shall omit the prefix ``chromo-'' when referring to field components.} field $\vec{E}$
 is the sum of a nonperturbative longitudinal field $\vec{E}^{\rm NP}$,   aligned with the interquark axis,  and a perturbative irrotational field, 
 $\vec{E}_{\rm C}$:
\begin{equation}
\label{1.3}
 \vec{E}(\vec{x}) \; = \; \vec{E}^{\rm NP}(\vec{x})     \;  + \;  \vec{E}_{\rm C}(\vec{x}) \; \; ,
\end{equation}
\begin{equation}
\label{1.4}
\vec{\nabla} \; \times \vec{E}_{\rm C}(\vec{x}) \; = \;  0     \;  \; .
\end{equation}
The field $\vec{E}^{\rm NP}$ is responsible for confinement and manifests itself as a smooth, spatially extended flux tube
  connecting the two static sources.
 The curl of the electric field defines a magnetic current $\vec{J}_{\rm M}$ that encircles the flux-tube axis and exhibits continuum scaling.
 After taking into account Eq.~(\ref{1.4}), one obtains
\begin{equation}
\label{1.5}
\vec{\nabla} \; \times \vec{E}(\vec{x}) \; = \;  \vec{\nabla} \; \times \vec{E}^{\rm NP}(\vec{x}) \; = \; \vec{J}_{\rm M}(\vec{x})     \;  \; .
\end{equation}
The current produces a Lorentz force density  $ \vec{f} \; = \; \vec{J}_{\rm M} \times \vec{E}^{\rm NP} $,
where both $\vec{J}_{\rm M}$ and $\vec{E}^{\rm NP}$ are determined directly from lattice simulations. This relation provides nontrivial support to the Maxwell picture of confinement, according to which confinement arises through the dual-superconductor mechanism, with electric flux tubes stabilized by circulating magnetic currents. In addition, theoretical analyses, such as those of Ref.~\cite{Cea:2023}, suggest that the longitudinal electric field is predominantly governed by its Abelian components, thereby  substantiating both the  Maxwell picture of confinement and   Eq.~(\ref{1.5}). \\
As it is widely believed,  in  presence of light quarks it is expected that the string between the static
quark-antiquark pair breaks at large enough distances, due to creation of a pair of light quarks
which recombine with the static quarks into two static-light mesons. 
String breaking  is one of the defining characteristics of a confining gauge theory with dynamical matter fields.
This kind of phenomenon is not accessible by perturbative QCD, it can only be examined by nonperturbative
methods such as lattice QCD. However,  this expected zero-temperature phenomenon has proven elusive in simulations 
of lattice QCD and could not be clearly observed in early lattice investigations~\cite{Laermann:1998gm,Kratochvila:2003zj}. 
In fact,  string breaking is usually interpreted  as a quantum-mechanical mixing phenomenon~\cite{Drummond:1998ar}. 
This means that the two states, the string state and the two meson
state, are both needed to describe the potential. After the string is broken, the meson state dominates the new ground state of this system.   
Near the critical separation  $d^*$, defined as the point where the Wilson loop and the static-light meson operator have equal overlap onto the ground state,
the two states mix. If there is mixing, both the ground state and first excited state are superpositions
of the string state and the two-meson state. The system undergoes a level crossing giving rise to an energy gap between the states.
Using this method, evidence for this kind of string breaking  was  first found in  Ref.~\cite{Bali:2005fu}  for $N_f$ = 2  QCD  with pion mass about 640 MeV.
The authors of Ref.~\cite{Bali:2005fu} reported a string-breaking distance $d^*$  $\simeq$ 1.248(13)  fm. 
A more recent analysis employed  $N_f$ = 2+1 flavors of  nonperturbatively improved dynamical Wilson fermions with  pion and kaon mass about  
 280 MeV and  460 MeV~\cite{Koch:2018puh,Bulava:2019iut}, reporting  a string-breaking distance $d^*$ $\approx$ 1.22 fm. 
 In Ref.~\cite{Bulava:2024jpj} an extrapolation of QCD with  $N_f$ = 2+1 flavors to the physical point led to  $d^* \; = \; $ 1.211(13) fm.
Note that this last estimate is slightly smaller that the one in Ref.~\cite{Bali:2005fu}, as expected since  the string-breaking distance should decrease 
with the sea quark mass. \\
The above discussion shows that the detection of string breaking, as well as the estimation of the associated string-breaking distance in QCD with dynamical quarks, necessarily relies on certain theoretical assumptions.
 It should be desirable to have a model independent criterion for elucidating string breaking in full QCD.
 In our previous paper~\cite{Baker:2024peg} we suggested that some  hints of the onset of string breaking could be
 obtained by looking at the nonperturbative electric field  as extracted {\it via} Eqs.~(\ref{1.3}) and (\ref{1.4}) from the connected
 correlation function,  Eq.~(\ref{connected1}). 
The main advantage  resides on the fact that we can look directly at the nonperturbative gauge invariant
longitudinal electric field, $\vec{E}^{\rm NP}$, in the region between two static sources, that is responsible for the formation of a well-defined confining 
flux tube characterized   by a nonzero effective string tension. Actually,
in Ref.~\cite{Baker:2024peg} it resulted that  the longitudinal nonperturbative electric field takes the shape of a flux tube whenever
the distance between the sources does not exceed a value of about 1.1 fm. For larger distances we found that  the transverse profile of the  longitudinal
nonperturbative electric field on the midplane between two sources  was strongly suppressed.
In other words, for distances above $\approx$ 1.1 fm we 
did not find evidences for the formation of the almost uniform flux tube between the two color static sources. 
 That analysis alone is certainly not enough to make any firm claim about the onset of string breaking, but it has provided some hints of its appearance, 
 to be corroborated by further tests.
The main aim of the present paper is to extend  and strengthen the analysis reported in Ref.~\cite{Baker:2024peg}, by employing 
  the two connected correlators  Eq.~(\ref{connected1}), where  the Schwinger line  is attached to the quark time line or to the antiquark time line. 
By combining the information obtained from the two operators, one can scan both the nonperturbative and perturbative electric fields over the entire region between the two static color sources.

While standard Euclidean lattice QCD remains the premier tool for evaluating 
string breaking, quantum computation offers a promising future approach to 
resolve real-time gauge dynamics. In fact,
the important phenomenon of string breaking has recently received renewed 
attention from quantum-computing and quantum-simulation approaches. 
Although based on different theoretical and computational frameworks, 
these studies~\cite{John:2026nonabelian,GonzalezCuadra:2025observation,De:2024observation,Ciavarella:2025string,Magnifico:2020realtime,Calajo:2024digital} share with our work the goal of accessing the flux-tube structure 
more directly, from its formation to its eventual breaking.
\\
The remainder of the paper is organized as follows. In Sect.~\ref{S2} we discuss our lattice setup and summarize the numerical outcomes. 
 The main results of the present paper on the  model-independent evidence of the string breaking and the ensuing estimate of the string-breaking
  distance are presented in Sect.~\ref{S3} that comprises three Subsections. Finally,  in Sect.~\ref{S4} we summarize the  
  results of the paper and draw our conclusions.
\section{Lattice setup and numerical results}
\label{S2}
\begin{table*}[th]
\begin{center} 
  \caption{Summary of the measurements.}
  \label{measurements}
\setlength{\tabcolsep}{17pt}
\begin{tabular}{cccrrcc}
\toprule
\midrule
\multicolumn{7}{c}{QCD (2+1) HISQ flavors ($m_l=1/27 m_s)$}\\
\midrule 
lattice             & $\beta=10/g^2$ & $a(\beta)$ [fm]   &  $T$ [\textrm{MeV}]  &  $d$ [lattice units]  &   $d$ [fm]      & no. of measurements     \\  \midrule
$ 48^4           $	&	6.880	     & 0.0963  &	43	 \hspace{0.2cm}            &    8   \hspace{0.6cm}             &   0.770       &   6000               \\
$ 48^4           $	&	6.880	     & 0.0963  &	43	 \hspace{0.2cm}            &   10   \hspace{0.6cm}             &   0.963       &   6000               \\
$ 48^4           $	&	6.880	     & 0.0963  &	43	 \hspace{0.2cm}            &   12   \hspace{0.6cm}             &   1.156       &  12000               \\
$ 48^4           $	&	6.880	     & 0.0963  &	43	 \hspace{0.2cm}            &   14   \hspace{0.6cm}             &   1.348       &  12000               \\ 
$ 48^4           $	&	6.832	     & 0.1009  &	41	 \hspace{0.2cm}            &   12   \hspace{0.6cm}             &   1.210       &  12000                 \\ \\
\toprule
\midrule
\multicolumn{7}{c}{QCD (2+1) HISQ flavors ($m_l=m_s)$} \\
\midrule
lattice             & $\beta=10/g^2$ & $a(\beta)$ [fm]   &  $T$ [\textrm{MeV}]  &  $d$ [lattice units]  &   $d$ [fm]      & no. of measurements     \\  \midrule

$ 24^4           $	&	6.472	     & 0.1445  &	57	 \hspace{0.2cm}            &   8   \hspace{0.6cm}             &   1.156       &  12000               \\ \\
\toprule
\midrule
\multicolumn{7}{c}{SU(3) pure gauge} \\
\midrule 
lattice             & $\beta=10/g^2$ & $a(\beta)$ [fm]   &  $T$ [\textrm{MeV}]  &  $d$ [lattice units]  &   $d$ [fm]      & no. of measurements     \\  \midrule
$ 24^4           $	&	5.861	     & 0.1205  &	68  \hspace{0.2cm}            &    8   \hspace{0.6cm}             &   0.964       &   6000               \\
$ 24^4           $	&	5.750	     & 0.1519  &	54  \hspace{0.2cm}            &    8   \hspace{0.6cm}             &   1.215       &  90000               \\
$ 48^4           $	&	5.904 	     & 0.1108  &	37  \hspace{0.2cm}            &   12   \hspace{0.6cm}             &   1.330       &  172000              \\

\bottomrule 
\end{tabular}
\end{center}
\end{table*}
\begin{figure*}[htbp]
\begin{center}
\includegraphics[width=0.8\linewidth,clip]{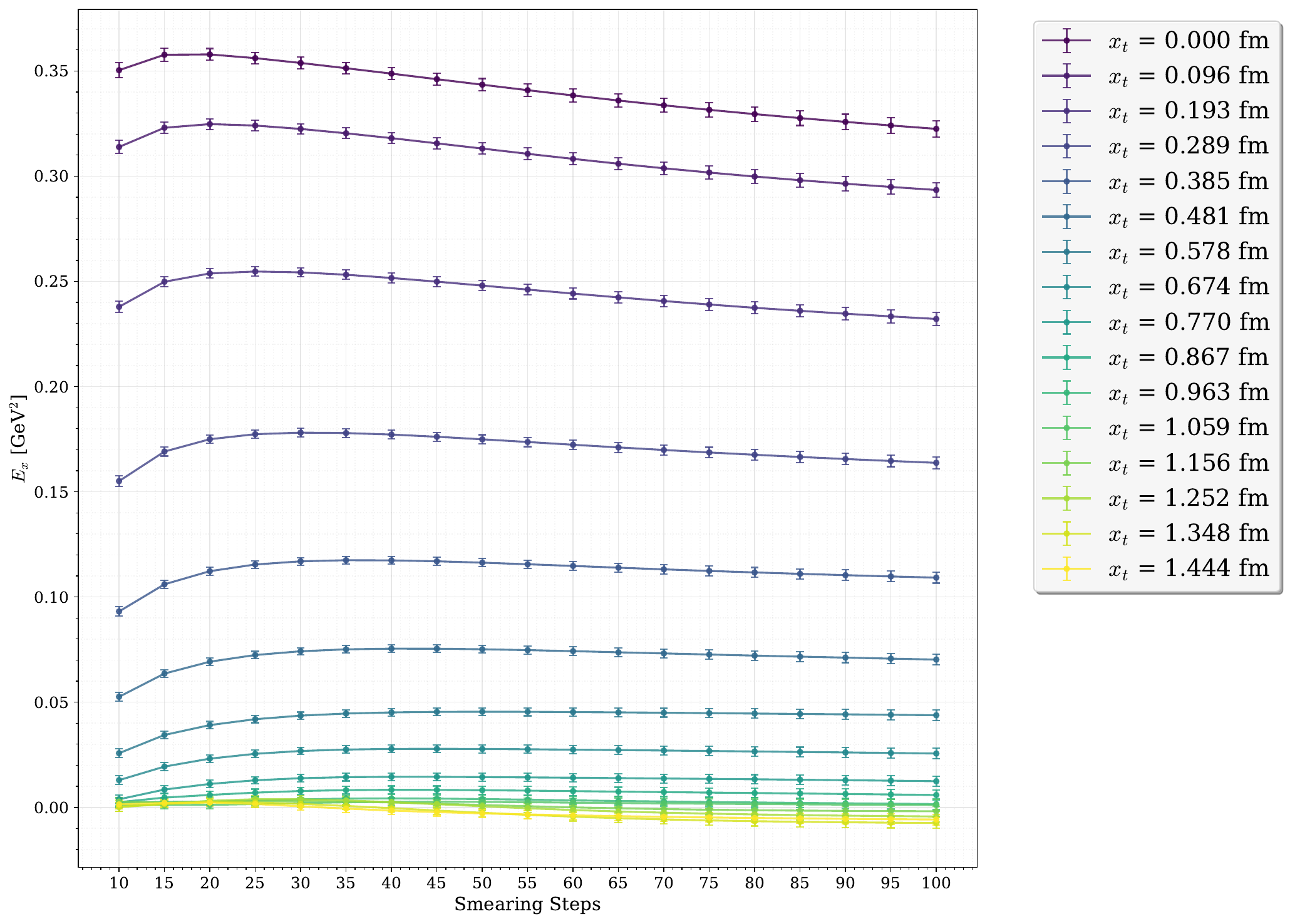}
\end{center}
\caption{Example of smearing. We consider here the case of QCD (2+1) HISQ flavors, $\beta=6.880$,  $d=10a \simeq 0.963$ fm, $E_x(x_l=3a,x_t)$.
}            
\label{Fig:smearing}
\end{figure*}
\begin{figure*}[htbp]
\begin{center}
\includegraphics[width=0.8\linewidth,clip]{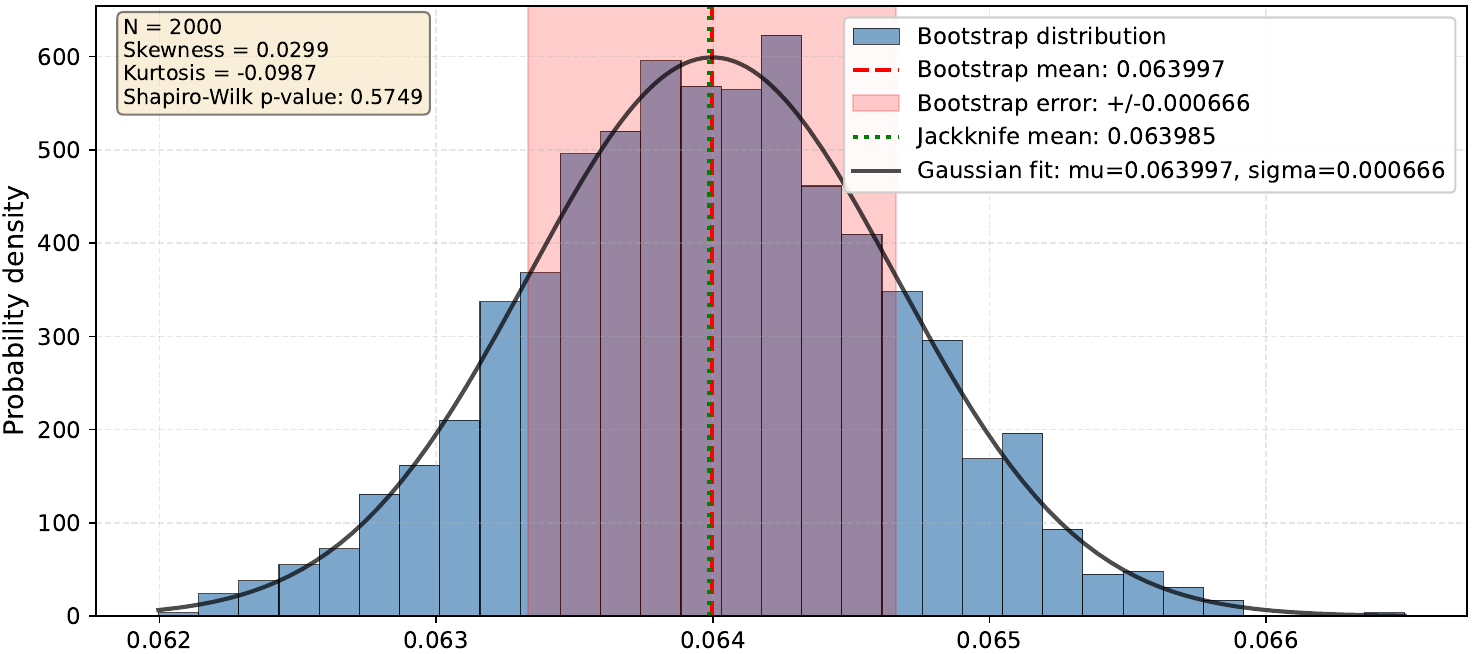}
\end{center}
\caption{Full bootstrap distribution of the raw data for $E_x(x_l=2a,x_t=a)$ on a $48^4$ lattice at $\beta=6.880$. The number of measurements is 6,000, the number of bootstrap samples considered is 2,000. The histogram and Gaussian overlay and confirm approximate normality — a check that the quoted  error is statistically meaningful for this ratio observable.
}            
\label{Fig:histogram}
\end{figure*}
We conduct lattice QCD simulations featuring $2+1$ flavors of Highly Improved Staggered Quarks (HISQ), employing the HISQ/tree action as detailed in Refs.~\cite{Follana:2006rc,Bazavov:2009bb,Bazavov:2010ru}. To ensure  that  our results remain physically relevant, we tune the couplings along a line of constant physics (LCP) following the methodology in Ref.~\cite{Bazavov:2011nk}. In this setup, the strange quark mass $m_s$ is set to its physical value, while the light-to-strange mass ratio is maintained at $m_l/m_s = 1/27$, which yields a physical pion mass of 140~MeV in the continuum limit.

Our simulations were performed on a $48^4$ lattice (see Table~1). Using the rational hybrid Monte Carlo (RHMC) algorithm, we generated thermalized configurations, saving one every 25 trajectories of unit length for subsequent analysis. These configurations served as the basis for measuring the electromagnetic field tensor in the presence of a static quark-antiquark pair at a fixed separation distance $d$. Finally, we set the lattice scale {\it via} the kaon decay constant $f_K$, following the prescription detailed in Appendix~B of Ref.~\cite{Bazavov:2011nk}.

The lattice spacing in physical units is given by~\cite{Bazavov:2011nk,MILC:2010hzw,Bazavov:2017dus}
\begin{equation}
\label{abeta}
a(\beta) \;=\; \frac{r_1}{r_1 f_K} \frac{c_0^K f(\beta)+c_2^K (10/\beta) f^3(\beta)}{
1+d_2^K (10/\beta) f^2(\beta)} \;,
\end{equation}
where
\begin{equation}
\label{r1fkappa}
\begin{split}
r_1 \;  &= \; 0.3106 \; \mathrm{fm} \;,
\\
 r_1 f_K \;  &= \; 
\frac{0.3106 \; \mathrm{fm} \; \cdot \; 156.1/\sqrt{2} \; \mathrm{MeV}}{
197.3 \; \mathrm{MeV \; fm}} \; \;
\end{split}
\end{equation}
and
 \begin{equation}
\label{coefficients}
c_0^K = 7.66 \; \;  , \; \;  c_2^K \; = \; 32911 \; \; , \; \;  d_2^K  \;  =  \;  2388 \; \; .
\end{equation}
In Eq.~(\ref{abeta})  $f(\beta)$ is the two-loop beta  function:
\begin{equation}
\label{2.6}
f(\beta)=[b_0 (10/\beta)]^{-b_1/(2 b_0^2)} \exp[-\beta/(20 b_0)] \; ,
\end{equation}
$b_0$ and $b_1$ being its universal coefficients,
\begin{equation}
\label{fsun1}
 b_0 \, = \, \frac{11}{(4\pi)^2} \; \; , \; \; b_1 \, = \, \frac{102}{(4\pi)^4} \; .
\end{equation} 
%

To investigate the potential impact of the strange quark mass on the string-breaking distance, we also performed $(2+1)$-flavor HISQ simulations for 
the unphysical symmetric case ($m_l = m_s$) on a $48^4$ lattice (see Table~\ref{measurements}); in this setup all pseudoscalar mesons are degenerate with a mass of about 700 MeV~\cite{Bazavov:2011nk}.

Since the infinite quark masses in the SU(3) pure-gauge  theory preclude the string-breaking phenomenon, a direct comparison with the behavior of SU(3) electric flux tubes is particularly instructive. We have measured the electric pure-gauge SU(3) fields at lattice distances both below and above the expected QCD string-breaking distance 
(see Table~\ref{measurements}).

In the case of SU(3) pure-gauge  theory we have used the standard Wilson action discretization  and set the physical scale for the lattice spacing according to Ref.~\cite{Necco:2001xg}:
\begin{equation}
\begin{gathered}
    a(\beta) = r_0 \exp \left[ c_0 + c_1(\beta-6) + c_2(\beta-6)^2 + c_3(\beta-6)^3 \right]\, \\
    r_0 = 0.5~\text{fm}, \quad c_0 = -1.6804, \quad c_1 = -1.7331, \\
    c_2 = 0.7849, \quad c_3 = -0.4428 \;,
\end{gathered}
\end{equation}
for all $\beta$ values in the range $5.7 \le \beta \le 6.92$. In this scheme, 
the value of the square root of the string tension is $\sqrt{\sigma} \approx 0.465$~GeV 
(see Eq.~(3.5) in Ref.~\cite{Necco:2001xg}).

Our setup employs a single step of four-dimensional hypercubic smearing (HYPt) on the temporal links, with parameters $(\alpha_1, \alpha_2, \alpha_3) = (1.0, 1.0, 0.5)$~\cite{Hasenfratz:2001hp}. Additionally, we apply $N_{\text{HYP3d}}$ steps of hypercubic smearing restricted to the spatial directions (HYP3d), using parameters $(\alpha_1^{\text{HYP3d}}, \alpha_3^{\text{HYP3d}}) = (0.75, 0.3)$.
\\
Figure~\ref{Fig:smearing} illustrates the typical evolution of the $E_x(x_l, x_t)$ component of the  field tensor under HYP3d smearing. The numerical results shown correspond to a $48^4$ lattice at $\beta=6.880$, with a longitudinal separation $x_l = 10a \simeq 0.963$~fm. As the figure demonstrates, 50 HYP3d smearing steps are sufficient to stabilize the results  for not too small transverse distance $x_t$.
For very small $x_t$ the data display a tiny drift of the central value with increasing smearing that, however, does not seem to be statistically significant.
Accordingly, the measurements reported in the remainder of this paper were performed after 50 HYP3d steps. We note that this smearing strategy differs from our previous work, where the number of smearing steps was tuned for each transverse coordinate $x_t$ to the "optimal"  value -- specifically, the point where the observable reaches its maximum. Anyway, we have verified that for the lattice setups investigated here, these two procedures do not yield significant differences. \\
In Table~\ref{measurements} we display the list of measurements we have done. We considered distances between the sources in a wide interval,
 ranging from 0.770~fm to  1.348~fm. We considered QCD with (2+1) flavour at  physical quark masses ($m_l = 1/27 m_s$) and 
 symmetric quark masses  ($m_l = m_s)$, as well as the SU(3) pure-gauge theory corresponding to quarks with infinite masses. \\
To estimate  the statistical  errors we used jackknife and bootstrap 
resampling methods. 
In Fig.~\ref{Fig:histogram} we present the full bootstrap distribution of the raw data 
corresponding to the longitudinal electric field $E_x(x_l=2a,x_t=a)$ for QCD (2+1) HISQ flavors on a  $48^4$ lattice at $\beta=6.880$, with distance 
$d = 10 \, a \simeq 0.963$ fm between the quark and the antiquark sources.
The number of measurements is 6,000 and the number of bootstrap samples considered is 2,000. 
The histogram and Gaussian overlay and confirm approximate normality, thus furnishing a check that the quoted values and  errors are statistically meaningful for our observables. 
\\
The electric field components, evaluated using the Schwin\-ger line attached to the quark time line or to the antiquark time line, have the following symmetry properties 
if there is a well defined flux-tube structure extending from the quark up to the antiquark:
\begin{eqnarray}
    E_x^{\rm quark}(x_l,x_t)&=&E_x^{\rm antiquark}(d-x_l,x_t) \;, \nonumber\\ 
    E_y^{\rm quark}(x_l,x_t)&=&-E_y^{\rm antiquark}(d-x_l,x_t) \;,\\
    E_z^{\rm quark}(x_l,x_t)&=&-E_z^{\rm antiquark}(d-x_l,x_t)\;.  \nonumber
\end{eqnarray}
Additionally, both $\vec{E}^{\rm quark}$ and $\vec{E}^{\rm antiquark}$ satisfy the following conditions as a consequence of cylindrical symmetry:
\begin{eqnarray}
\label{antisym1}
E_x(x_l, x_t) & = & E_x(x_l,-x_t) \;, \nonumber\\ 
    E_y(x_l, x_t) & = & - E_y(x_l, -x_t) \;,\\
    E_z(x_l, x_t) & = & E_z(x_l, x_t)\;.  \nonumber
\end{eqnarray}
\begin{figure*}[htbp]
    \centering
    \begin{subfigure}[b]{0.32\textwidth}
        \centering
        \includegraphics[width=\textwidth]{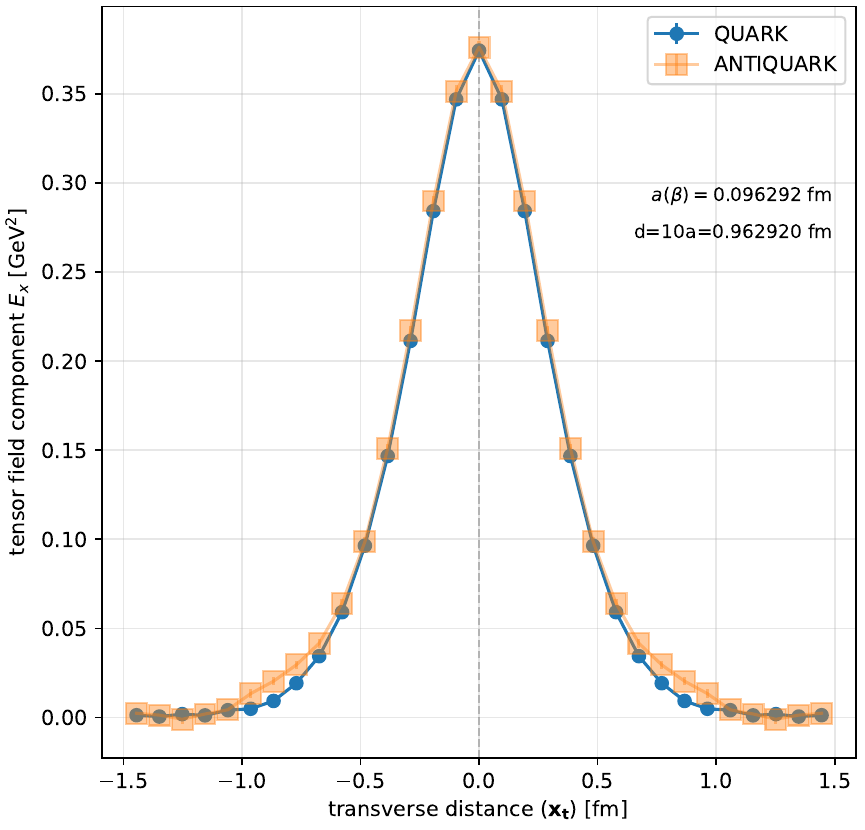}
        \caption{$E_x(x_l=4a,x_t)$}
        \label{fig:plot1}
    \end{subfigure}
    \hfill
    \begin{subfigure}[b]{0.32\textwidth}
        \centering
        \includegraphics[width=\textwidth]{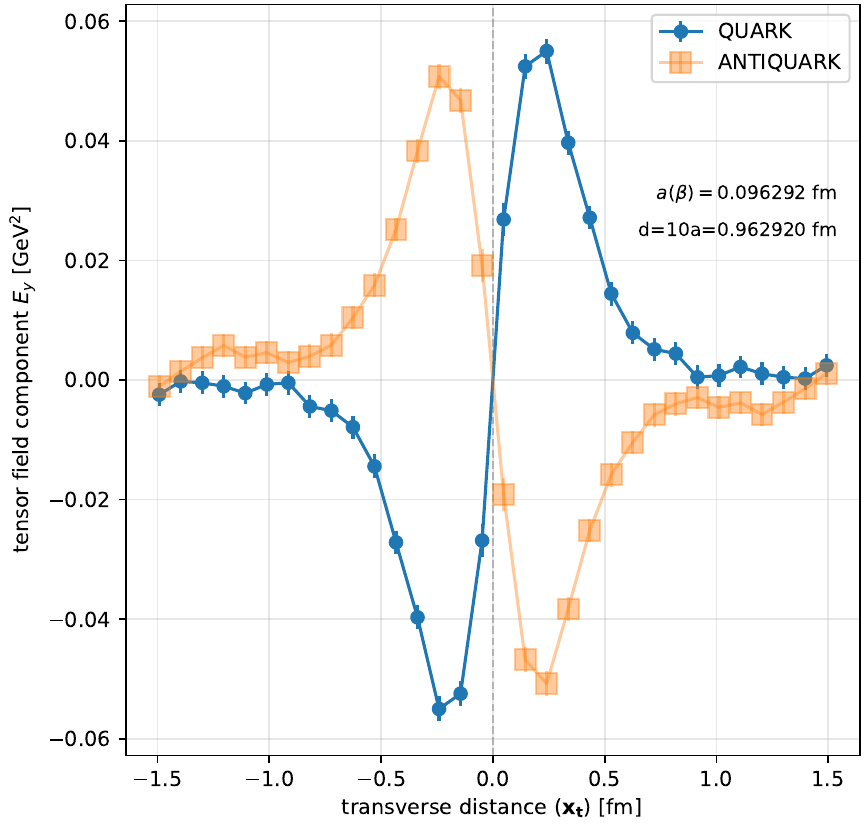}
        \caption{$E_y(x_l=3a,x_t)$}
        \label{fig:plot2}
    \end{subfigure}
    \hfill
    \begin{subfigure}[b]{0.32\textwidth}
        \centering
        \includegraphics[width=\textwidth]{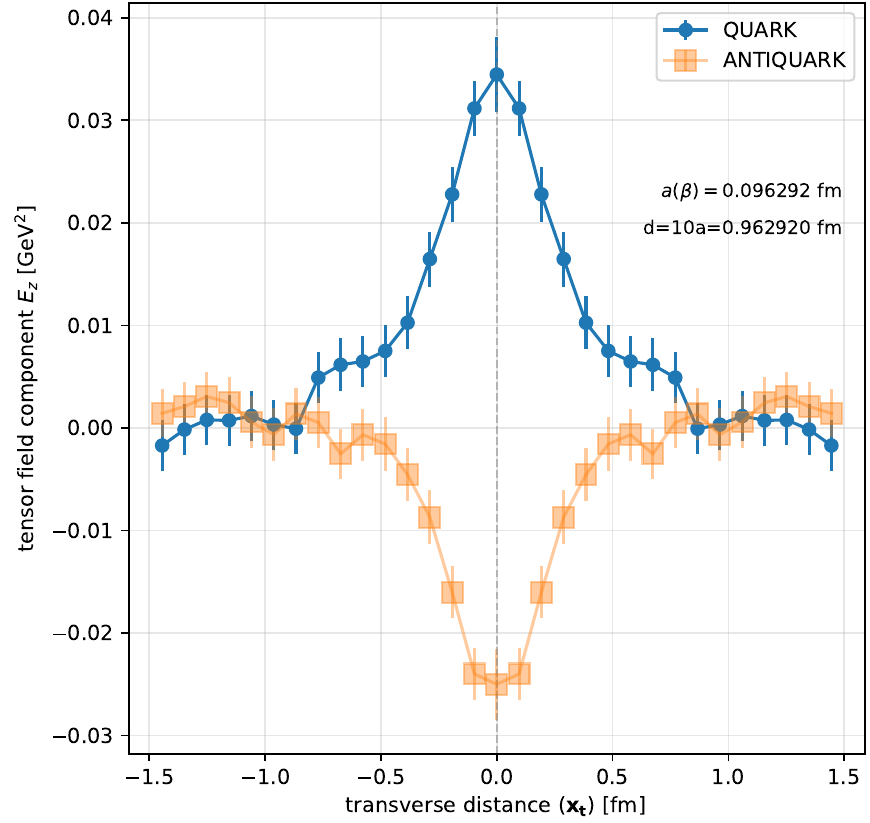}
        \caption{$E_z(x_l=3a,x_t)$}
        \label{fig:plot3}
    \end{subfigure}
    \caption{A sample of the $E_x$, $E_y$, $E_z$ fields obtained from the ``quark correlator'' (Schwinger line attached to the quark time line in the Wilson loop) and the ``antiquark correlator'' (Schwinger line attached to the antiquark time line in the Wilson loop). The data refer to the case of QCD (2+1) HISQ flavors, $48^4$ lattice, $\beta=6.880$, with distance $d=10a \simeq 0.963$ fm between the quark and the antiquark, and 50 HYP3d smearing steps.}
    \label{Fig:Exyz}
\end{figure*}
Indeed, in  Fig.~\ref{Fig:Exyz} we report the longitudinal and transverse electric fields  obtained from the connected  correlators with  the Schwinger line attached 
to the quark time line and  to the antiquark time line in the Wilson loop corresponding to  QCD  with (2+1) HISQ flavours on a $48^4$ lattice at $\beta = 6.880$, with distance between the static color sources $d=10a=0.963$ fm.  We see that the data within the statistical uncertainness  satisfy almost perfectly the expected symmetry or 
antisymmetry properties.
\begin{figure}[htbp]
\begin{center}
\includegraphics[width=0.95\linewidth,clip]{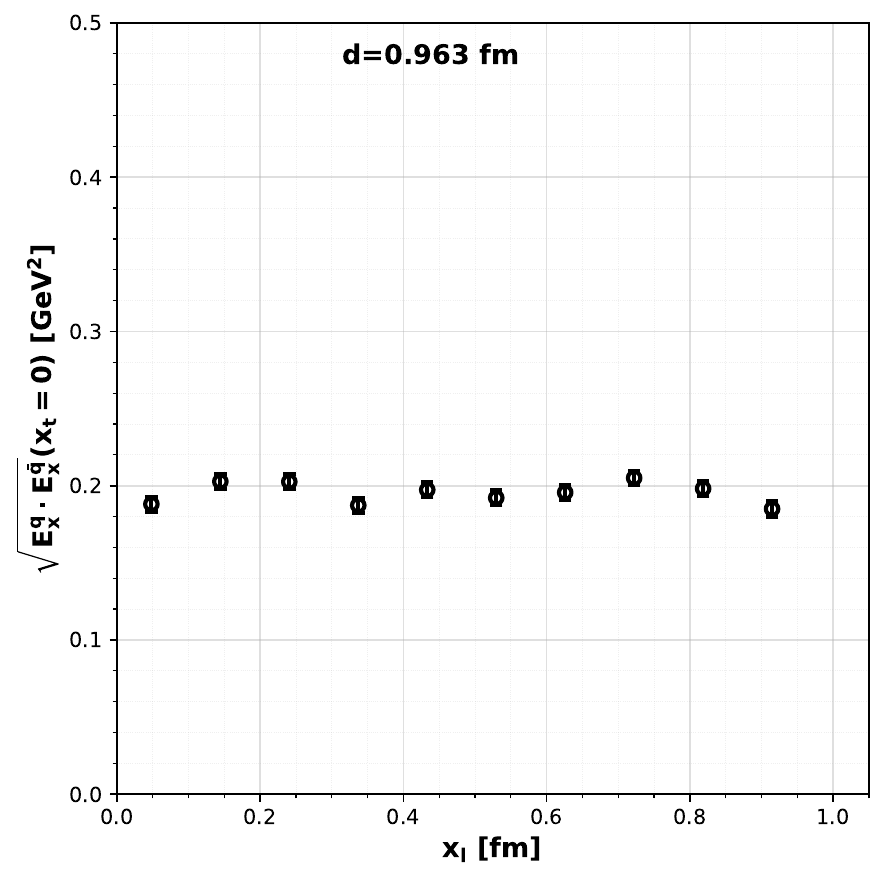}
\end{center}
\caption{The quantity $\sqrt{E_{x,q}^{\rm NP}(x_l,x_t=0) E_{x,\bar{q}}^{\rm NP}(x_l,x_t=0)}$ in the case of QCD (2+1) HISQ flavors, $48^4$ lattice, $\beta=6.880$, $d=10a \simeq 0.963$ fm.
}            
\label{Fig:Exqqbar}
\end{figure}

In Fig.~\ref{Fig:Exqqbar}, we provide a plot of $\sqrt{E_{x,q}^{\rm NP}(x_l,x_t) E_{x,\bar{q}}^{\rm NP}(x_l,x_t)}$ 
at $x_t = 0$. The motivation for choosing this quantity lies in the cancellation of the multiplicative renormalization factors 
that appear in the lattice evaluation of the chromoelectric fields: the renormalized field value $E_{x,q}^{\rm NP}(x_l,x_t)$ 
gets a renormalization factor $A^{x_l + x_t}$, where $A$ is a constant independent on $x_l$, $x_t$, and $x_l + x_t$ is the length of the Schwinger line for the quark operator, and correspondingly the antiquark operator gets a renormalization factor $A^{d - x_l + x_t}$.
In this way the renormalization factor in $\sqrt{E_{x,q}^{\rm NP}(x_l,x_t) E_{x,\bar{q}}^{\rm NP}(x_l,x_t)}$ is equal to $A^{d/2 + x_t}$ -- 
independent from $x_l$. While smearing in our approach acts as an effective renormalization, the use of finite smearing might result in an effective value of the factor $A$ close, but not exactly equal to 1 -- which would be more visible on observables with large Schwinger line length. The fact that the field values in Fig.~\ref{Fig:Exqqbar} are largely independent from $x_l$ supports the assumption that 
the fully renormalized nonperturbative longitudinal field does not depend on $x_l$ when the flux tube is present, and, in particular, the asymmetry between the field close to the quark, and the field close to the antiquark comes from this renormalization issue, and does not signify a physical asymmetry. 

Finally, we have also verified that our lattice setup is sufficiently close to the continuum limit by checking that different choices of lattice parameters, corresponding 
to the same physical  distance \(d\) between the sources, yield consistent values of the relevant observables when expressed in physical units.  
\section{String breaking}
\label{S3}
\begin{figure*}[htbp]
\begin{center}
\includegraphics[width=0.49\linewidth,clip]{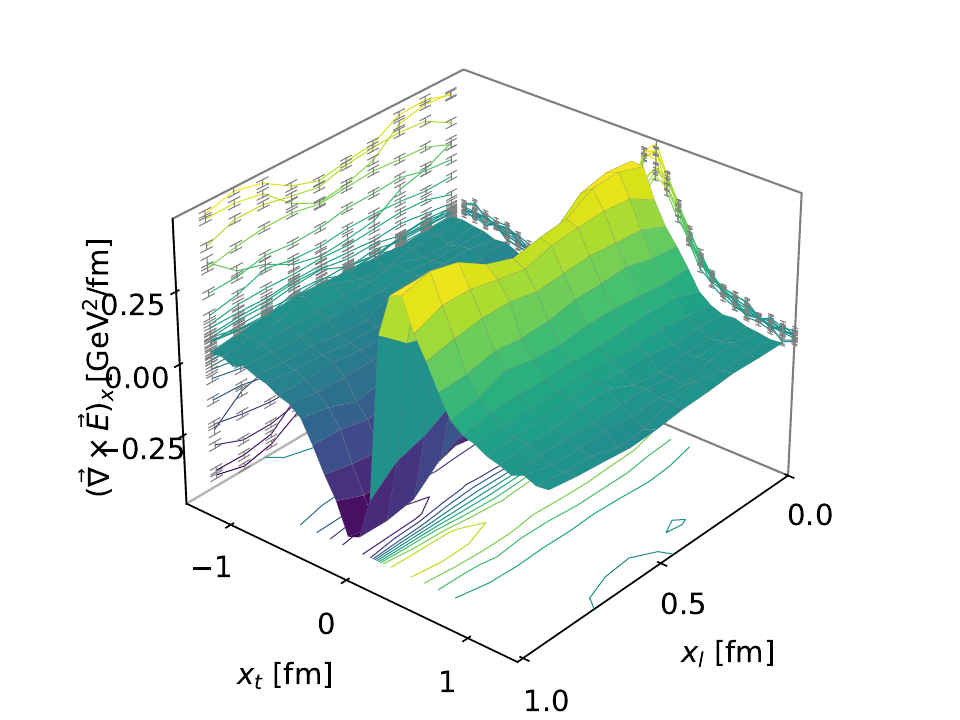}
\includegraphics[width=0.49\linewidth,clip]{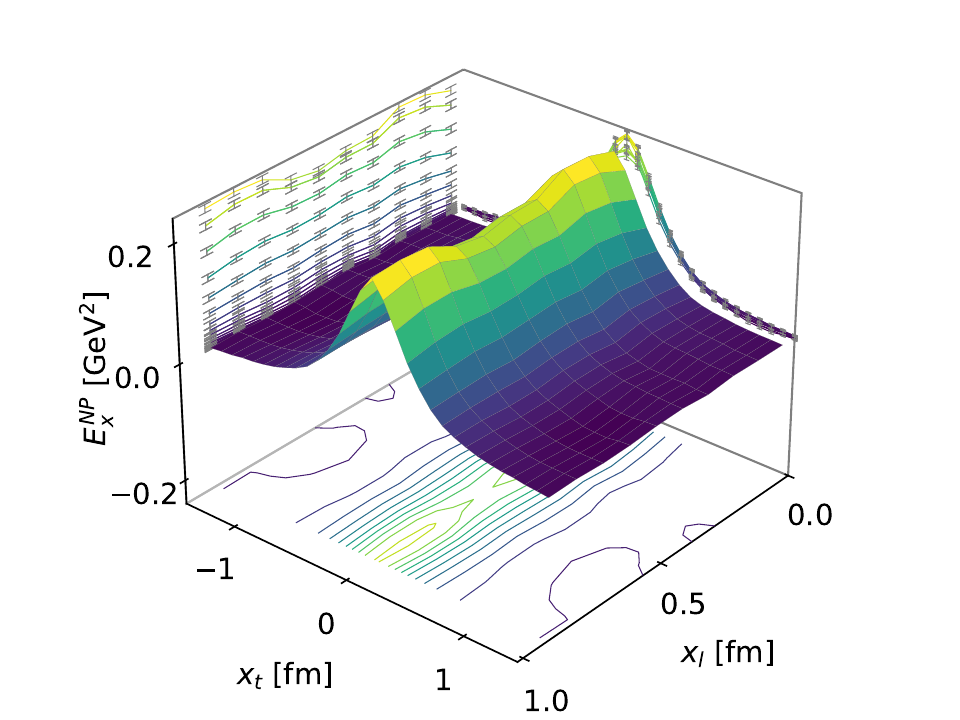}
\end{center}
\caption{The full 3D profile of  the magnetic current and the longitudinal nonperturbative electric  field 
for QCD$_{2+1}$   with $m_{\pi} \simeq $ 140 MeV at $\beta=6.880$ and $d=10 \, a \simeq 0.963$ fm.
}            
\label{Fig3.1}
\end{figure*}
 The main goal of this paper  is to investigate  the   possible breaking of the flux-tube structure generated by 
 a static quark-antiquark pair in QCD with 2+1 HISQ  flavors at the physical point. As we said before,
 we tuned the light and strange  quark masses  along the line of constant physics   corresponding to a physical pion mass of $m_{\pi} \simeq  140$ MeV. 
 From the results of our previous works we know that  the fingerprint of quark confinement is the presence of a squeezed flux-tube made of longitudinal
 electric field almost uniform along the line joining  the quark-antiquark static color charges. Indeed, it is easy to see  that such a color-field  structure gives rise
 to confining  potential increasing linearly with the distance $d$ between the static  charges. 
The main advantage of the present  work resides on the fact that we can look directly at  both the 
nonperturbative   electric field $\vec E^\text{NP}$ and  the perturbative irrotational field $\vec{E}_{\rm C}$, expected
to be the dominant one near the quark and antiquark color sources, along the whole region between the quark-antiquark pair.
In particular,  we shall employ  the connected correlation functions where  the Schwinger line is attached to the quark time  line 
(quark connected correlator) or to the antiquark time line (antiquark connected correlator). In this way, if the string does not break, there is  an almost perfect
quark-antiquark symmetry for the longitudinal electric fields, $E_x$,  and antisymmetry for the transverse electric fields, $E_y$ and $E_z$ 
 -- see~Eq.~(\ref{antisym1}). 
We have already checked~\cite{Baker:2025bja} that on the midplane of the static color sources both quark and antiquark connected correlators returned electric fields  consistent within the  statistical errors. 
Our strategy is to consider quark-antiquark static color sources separated by a given distance $d$ and look at both the magnetic current and the longitudinal
nonperturbative electric field $\vec E^{\rm NP}$. In fact, if there are almost uniform magnetic current and nonperturbative electric field along the line
joining  the quark-antiquark color charges, then we can affirm that there is a uniform flux tube without any string breaking. In this case we should
observe an almost perfect quark-antiquark symmetry and antisymmetry for the longitudinal and transverse electric fields, respectively.
By increasing the quark-antiquark distance $d$, one expects that the string breaks at some point within the flux tube. 
In this case  we should  observe a drastic drop in the nonperturbative electric field $\vec E^{\rm NP}$ within the region between the two color static sources, thus giving  evidence for the breaking of the flux tube. In addition, the string breaking should also manifest itself as a loss of the quark-antiquark
symmetry or antisymmetry for the electric fields. 
\subsection{Below the string breaking distance}
\label{S3.1}
\begin{figure}
\includegraphics[width=0.47\textwidth,clip]{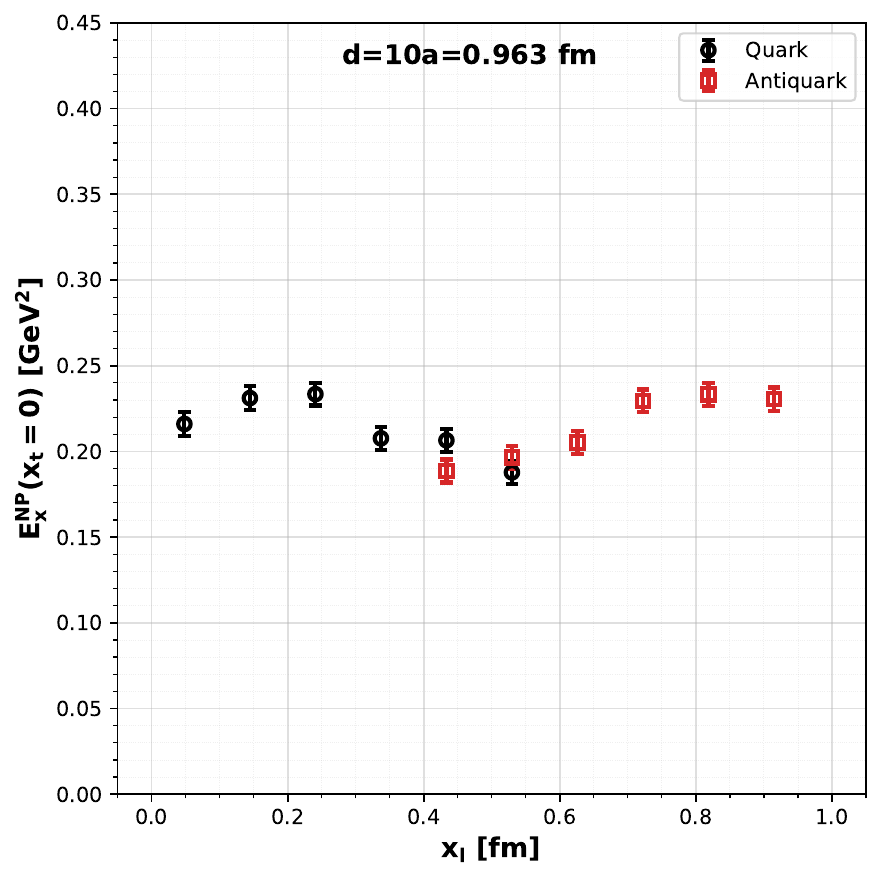}
\caption{\label{Fig3.2}  The nonperturbative electric field at $x_t = 0$ along the flux tube for $d=10 \, a \simeq 0.963$ fm.}
\end{figure}

\begin{figure*}
\begin{center}
\includegraphics[width=0.48\textwidth,clip]{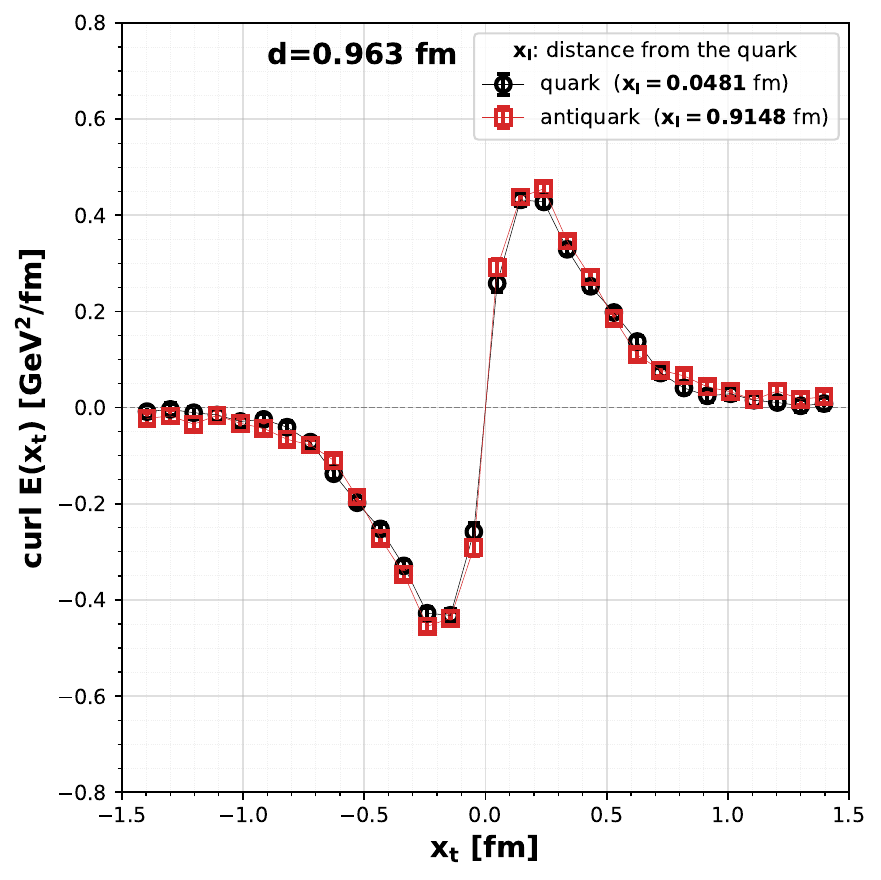}
\includegraphics[width=0.48\textwidth,clip]{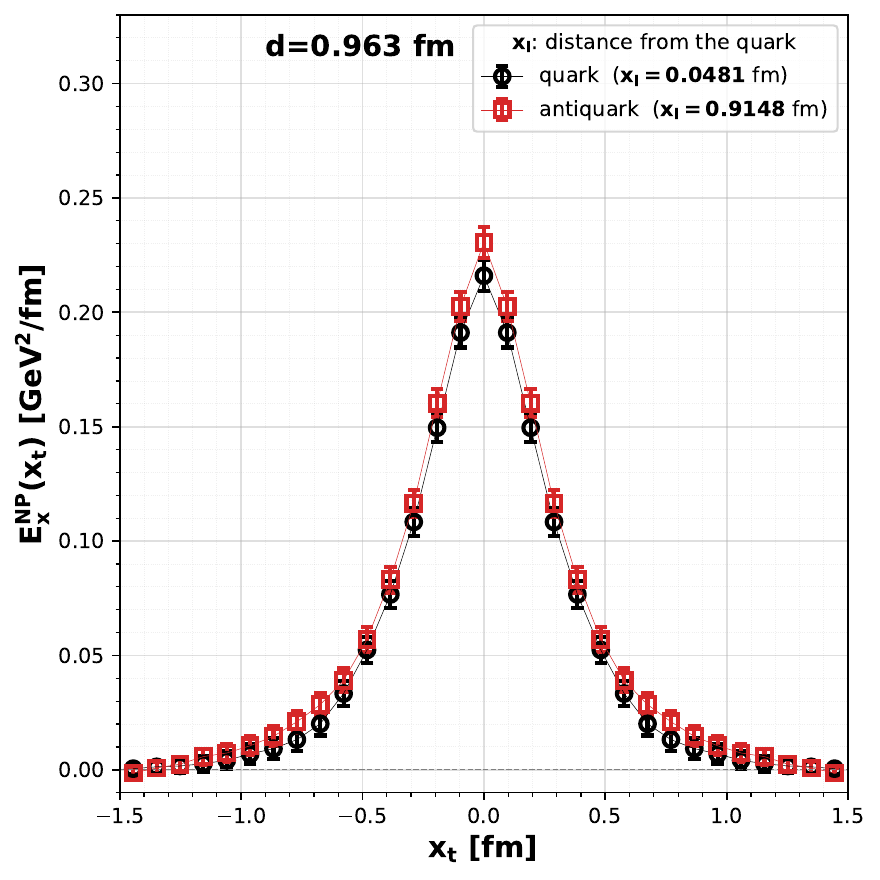}
\\
\includegraphics[width=0.47\textwidth,clip]{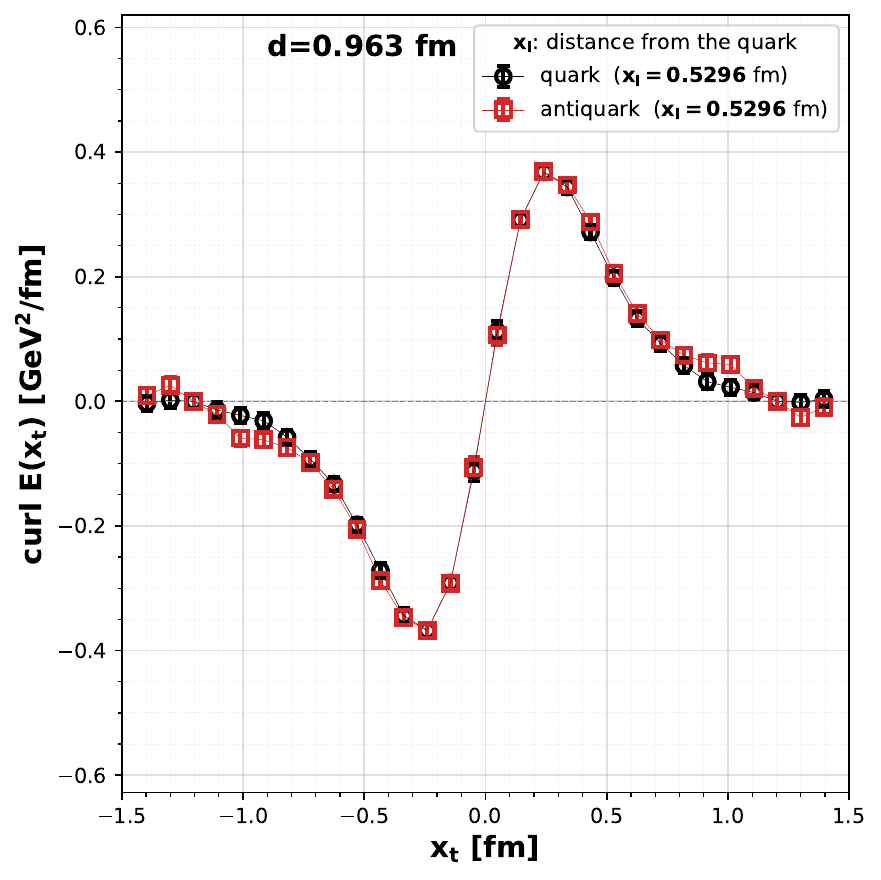}
\includegraphics[width=0.47\textwidth,clip]{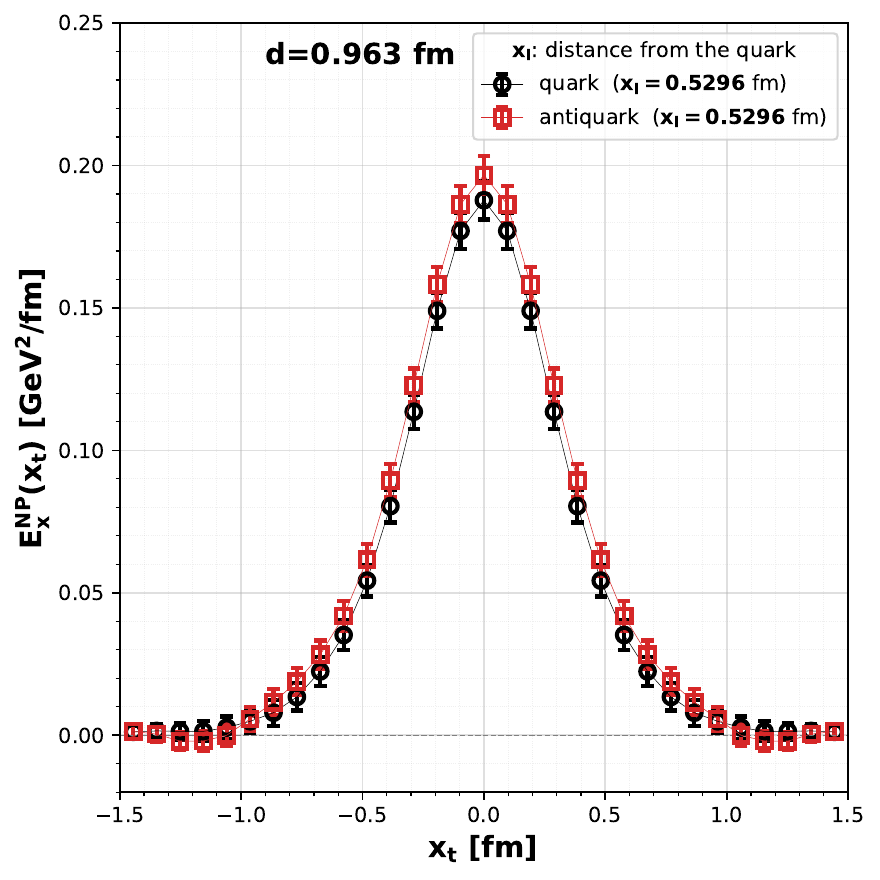}
%
\caption{\label{Fig3.3} Transverse distribution of the magnetic current and nonperturbative electric field
near the color sources and at the middle point for $d=10 \, a \simeq 0.963$ fm. }
\end{center}
\end{figure*}

\begin{figure*}
\begin{center}
\includegraphics[width=0.47\textwidth,clip]{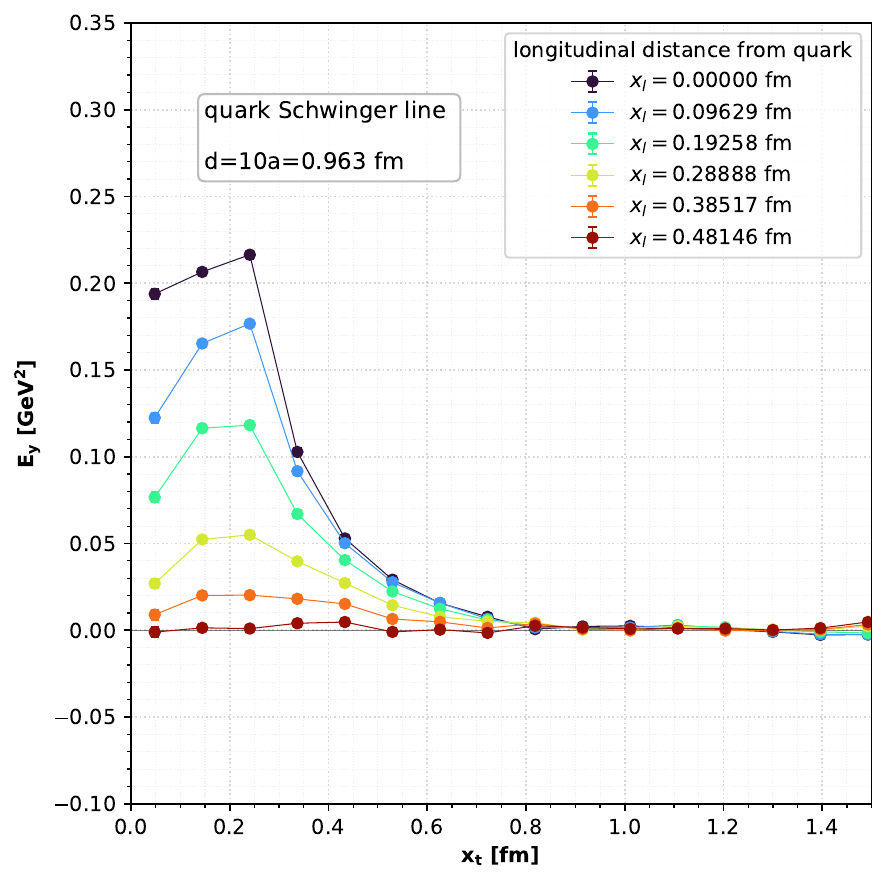}
\includegraphics[width=0.47\textwidth,clip]{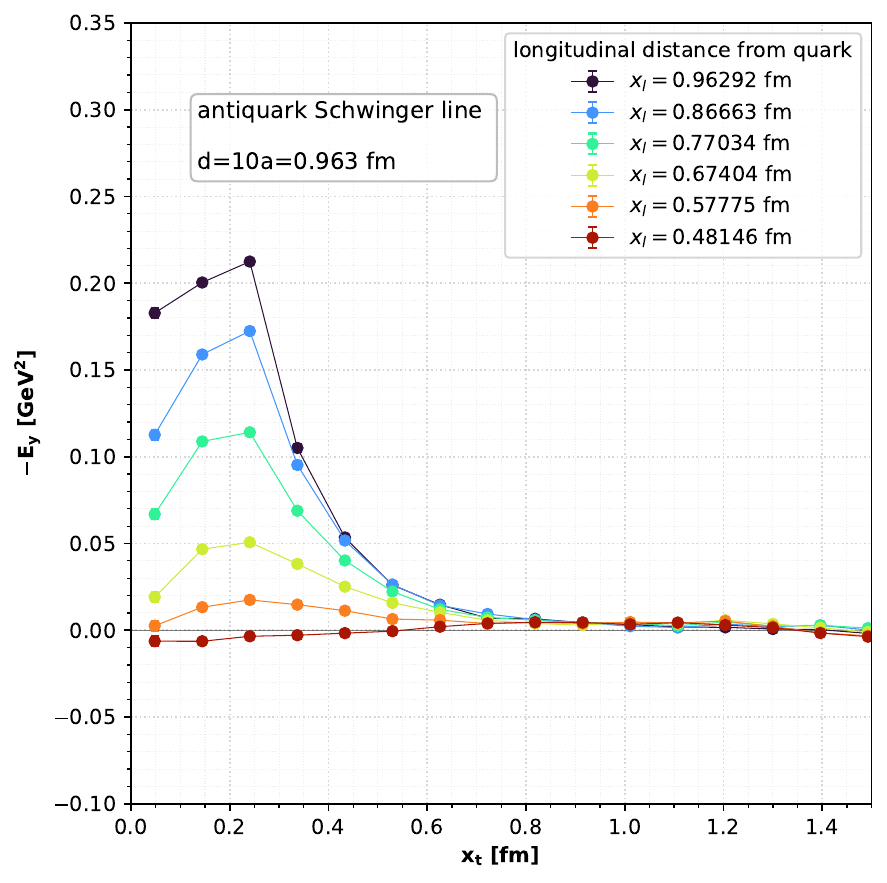}
\\
\includegraphics[width=0.47\textwidth,clip]{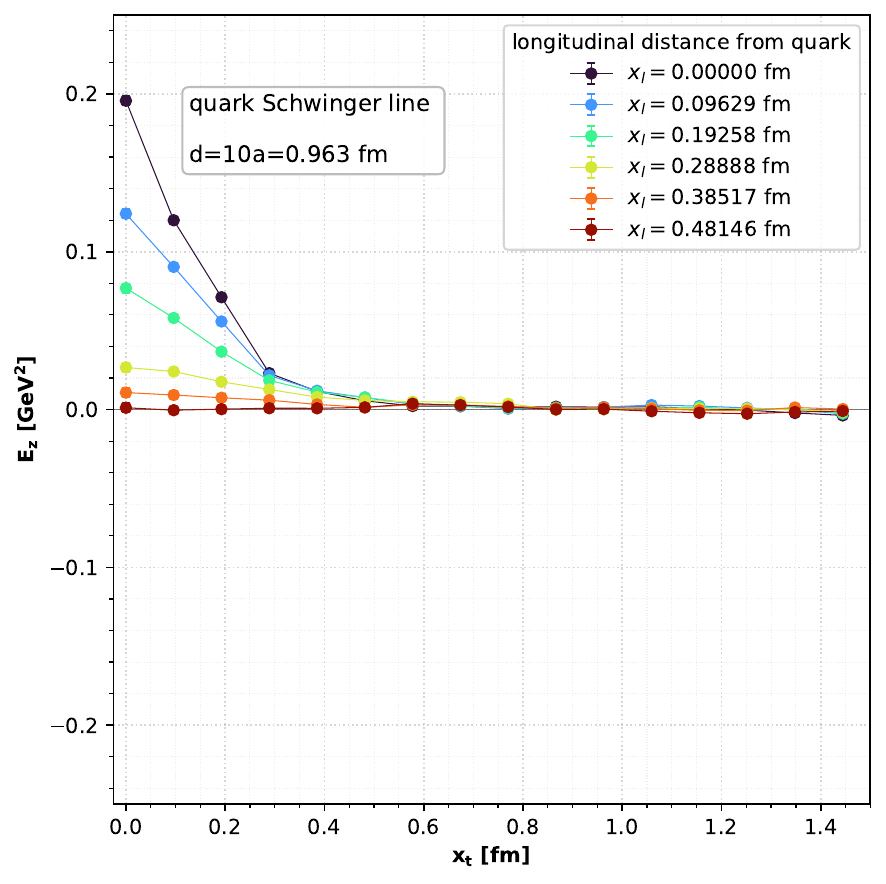}
\includegraphics[width=0.47\textwidth,clip]{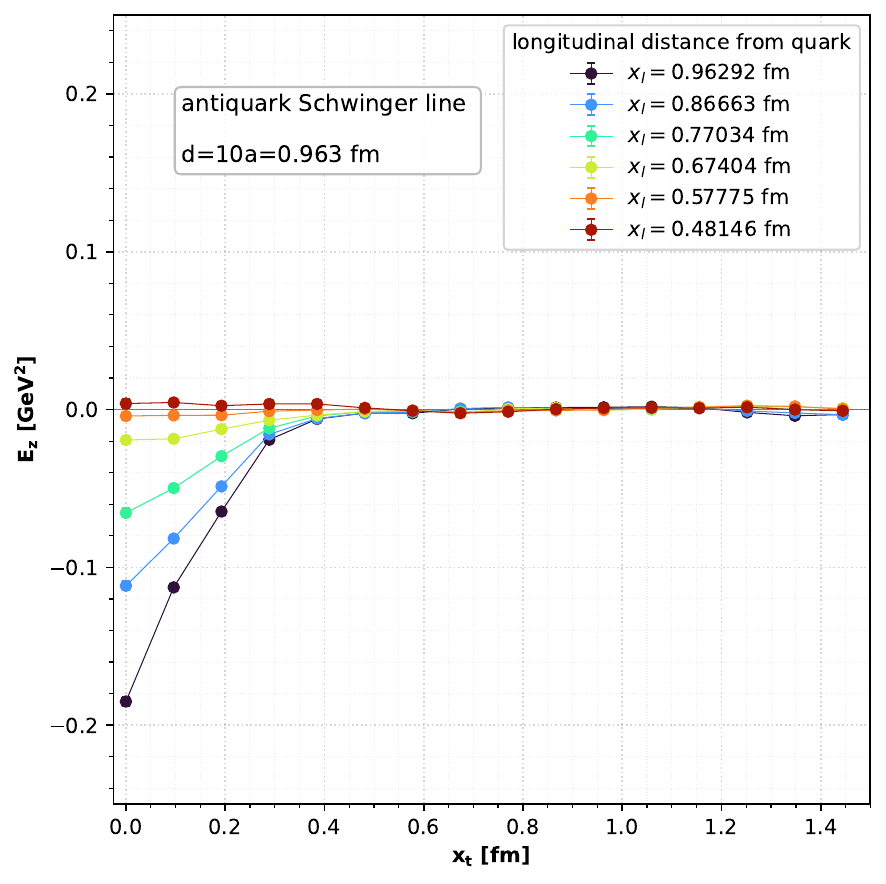}
%
\caption{\label{Fig3.4} Transverse distribution of the electric fields $E_y$ and $E_z$
along the flux tube for $d=10 \, a \simeq 0.963$ fm. }
\end{center}
\end{figure*}

\begin{figure}
\begin{center}
\includegraphics[width=0.47\textwidth,clip]{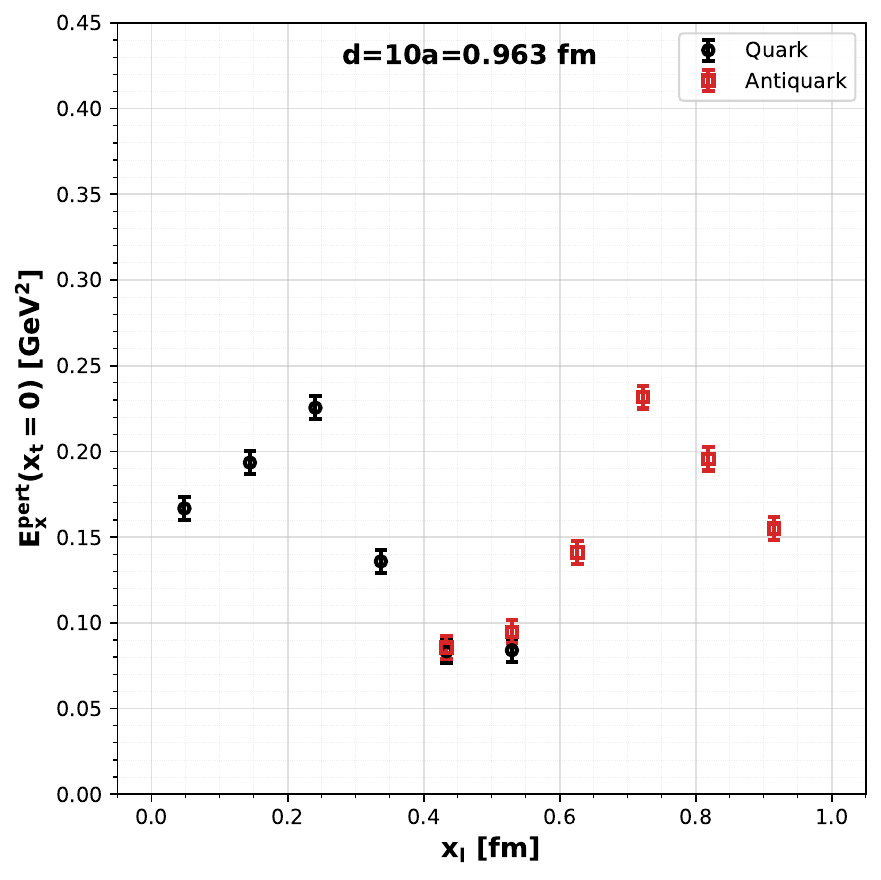}
\\
\includegraphics[width=0.47\textwidth,clip]{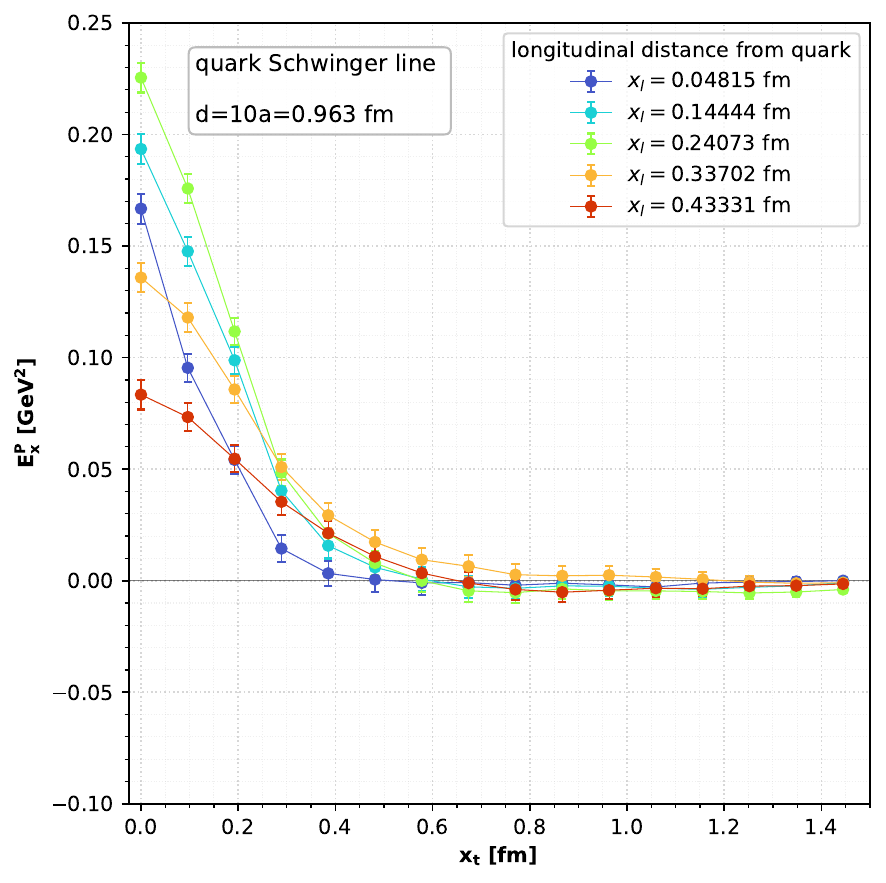}
\\
\includegraphics[width=0.47\textwidth,clip]{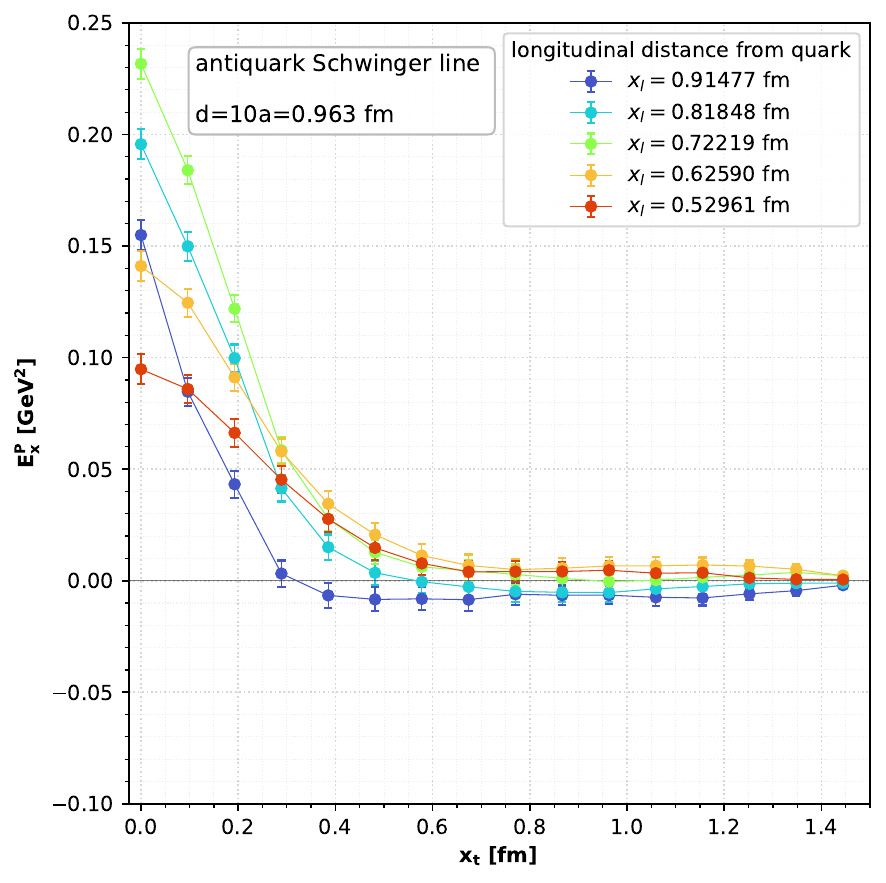}
%
\caption{\label{Fig3.5} The perturbative longitudinal electric field $E^P_x$ at $x_t = 0$ {\it versus} the longitudinal distance $x_l$ (upper panel).
Transverse distribution of   $E^P_x$  along the flux tube  for $d=10 \, a \simeq 0.963$ fm (middle and lower panels). }
\end{center}
\end{figure}

We begin by displaying in Fig.~\ref{Fig3.1}   the full three-dimensional plots of the magnetic current as defined by  Eq.~(\ref{1.5}) and the
nonperurbative longitudinal electric field $E_x^{\rm NP}$ for QCD with 
$(2+1)$-flavour  with quark masses at the physical values    at $\beta=6.880$ and
distance between the static quark-antiquark charges  $d=10 \, a \simeq 0.963$ fm. In this figure we show the symmetrized version of the flux-tube profile:
\begin{equation}
    \label{symmetriprofile1}
    J_{\rm M}(x_l,x_t) = 
\begin{cases}
J_{\rm M}^{\rm quark}(x_l,x_t) & \text{if } \; x_l < d/2, \\
J_{\rm M}^{\rm antiquark}(d-x_l,x_t) & \text{if } \;  x_l \ge d/2 \; ,
\end{cases}
\end{equation}
\begin{equation}
    \label{symmetriprofile2}
    E_x^{\rm NP}(x_l,x_t) = 
\begin{cases}
E_x^{\rm NP,\,quark}(x_l,x_t) & \text{if } \; x_l < d/2, \\
E_x^{\rm NP,\,antiquark}(d-x_l,x_t) & \text{if } \; x_l \ge d/2 \; .
\end{cases}
\end{equation}
Looking at Fig.~\ref{Fig3.1} we clearly see that the transverse shape of both the magnetic current and nonperturbative electric field does not 
depend on the longitudinal coordinate $x_l$, unambiguously signalling the presence of a well-defined  flux-tube structure.
For a more quantitative analysis, in Fig.~\ref{Fig3.2}  we show the profile of the nonperturbative longitudinal electric field  $E_x^{\rm NP}(x_l,x_t=0)$
along the flux tube. As a rule, there and in the following, for $0 \le  x_l  \le \frac{d}{2}$ the data points have been obtained from the quark connected correlator, 
while for $\frac{d}{2}   \le  x_l  \le  d$ they come
from the antiquark connected correlator. Note that at the middle point the two different estimates of the
 nonperturbative longitudinal electric field are in nice  agreement. \\
Figure~\ref{Fig3.2}  makes manifest the uniform behavior of the nonperturbative longitudinal electric field leading to an almost constant string tension along the whole flux tube. Moreover, it is evident that  $E_x^{\rm NP}(x_l,x_t)$ satisfies  the expected quark-antiquark symmetry.
This is  supported more quantitatively in Fig.~\ref{Fig3.3}, where the transverse distributions of the magnetic current and  the  nonperturbative electric field are displayed near the static color sources and at the middle point.  In both cases the
transverse distributions are in quite good agreement, fully supporting the quark-antiquark symmetry for both the magnetic current and  the longitudinal nonperturbative 
electric field. \\
The presence of a well-defined flux tube between the static color charges, leading to the expected quark-antiquark
symmetries, is further supported by looking at the transverse electric field $E_y$ and $E_z$. 
The electric fields $E_y$ and $E_z$ should be the transverse component of the  perturbative irrotational field  $\vec{E}_{\rm C}$.
This follows from Eq.~(\ref{1.3}) and the fact that  $\vec{E}^{\rm NP}$ is longitudinal, {\it i.e.} directed along the flux tube.
The longitudinal component of the perturbative irrotational field, that will be indicated as $E^P_x$,  can be found by solving numerically
the equation $(\vec{\nabla} \; \times \vec{E})_z  \; = \;  0 $.  So that, we see that $(E^P_x,E_y,E_z)$ can be thought as the three components of  $\vec{E}_{\rm C}$.  From the physical point of view, the perturbative electric field $\vec{E}_{\rm C}$ must be identified with
the Coulomb electric field generated by the quark-antiquark static color charges. If this is the case, the transverse electric fields
$E_y$ and $E_z$ are quark-antiquark antisymmetric, while  $E^P_x$ must be symmetric. Indeed, looking at Fig.~\ref{Fig3.4}, we see
that the transverse electric fields are sizable near the static color sources and satisfy an almost perfect quark-antiquark antisymmetry.
On the other hand, the longitudinal perturbative electric field should behave symmetrically under the quark-antiquark interchange.
In fact,  Fig.~\ref{Fig3.5}, displaying the profile along the flux tube of  $E^P_x$ at $x_t = 0$ together with the transverse distributions along the
flux tube of the longitudinal component of the perturbative electric field, shows clearly the expected quark-antiquark symmetry. \\
We already pointed out that, from the physical point of view, the perturbative electric field should be the Coulomb field that originates from
the static quark-antiquark charges. Remarkably, this is fully supported by the lattice data.  Indeed,
the transverse fields $E_{y}$ and $E_{z}$ behave as the transverse component of the Coulomb field,
{\it i.e.} they satisfy the  quark-antiquark antisymmetry, being positive in the first half of the flux tube and negative
in the second half. In addition the transverse field can be well fitted by a screened Coulomb field
generated by a static quark-antiquark pair. It turns out that one can obtain satisfying fits  with only   two parameters, namely
 the charge $Q$ and the screening mass $\mu$.
The longitudinal perturbative field  respects perfectly the quark-antiquark  symmetry and it is strictly positive along the flux tube,
as expected if  $E^P_x(\vec{x})$  is the longitudinal component of the Coulomb field generated by a static   quark-antiquark pair.  
Even more,  $E^P_x$ can be successfully fitted by the longitudinal screened Coulomb field due to a static  quark-antiquark pair with fitting 
parameters quite consistent with the ones obtained  from the fits of the transverse fields $E_{y}$ and $E_{z}$. Anyway,
a full account on this matter goes beyond the aim of the present paper. We plan to present  a rather complete discussion in a 
forthcoming paper. 
\subsection{Above the string-breaking distance}
\label{S3.2}
\begin{figure}
\includegraphics[width=0.47\textwidth,clip]{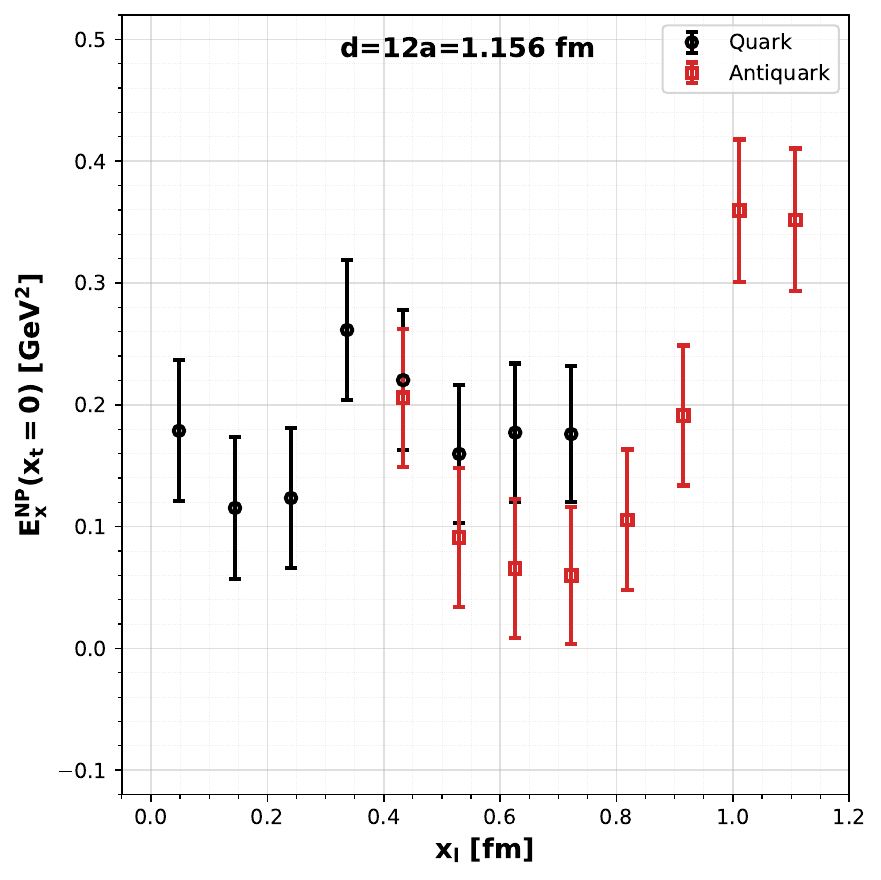}
%
\caption{\label{Fig3.6}  The nonperturbative electric field at $x_t = 0$ along the flux tube for $d =12 \, a  \simeq  1.156 $ fm.}
\end{figure}
\begin{figure*}
\begin{center}
\includegraphics[width=0.47\textwidth,clip]{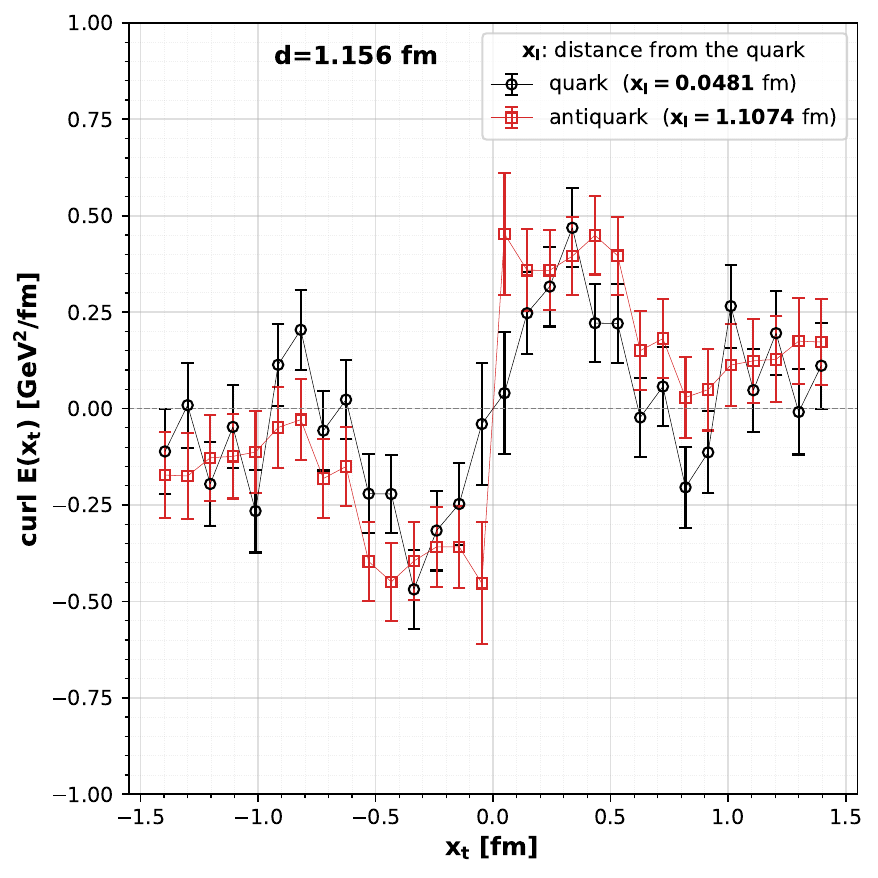}
\includegraphics[width=0.47\textwidth,clip]{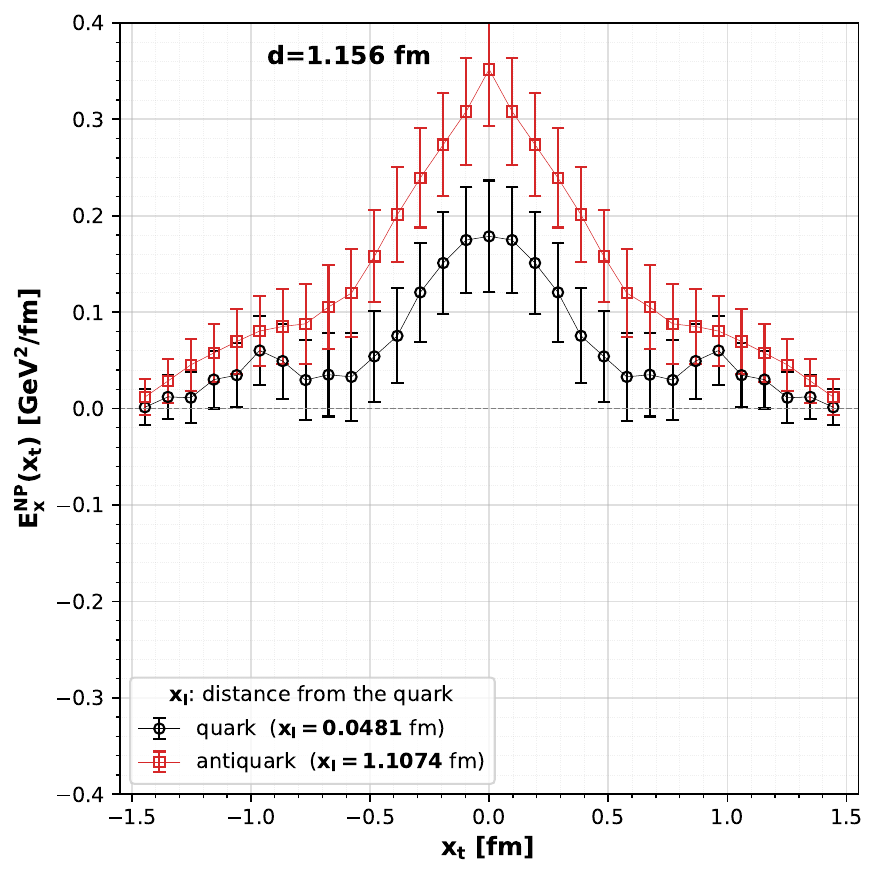}
\\
\includegraphics[width=0.47\textwidth,clip]{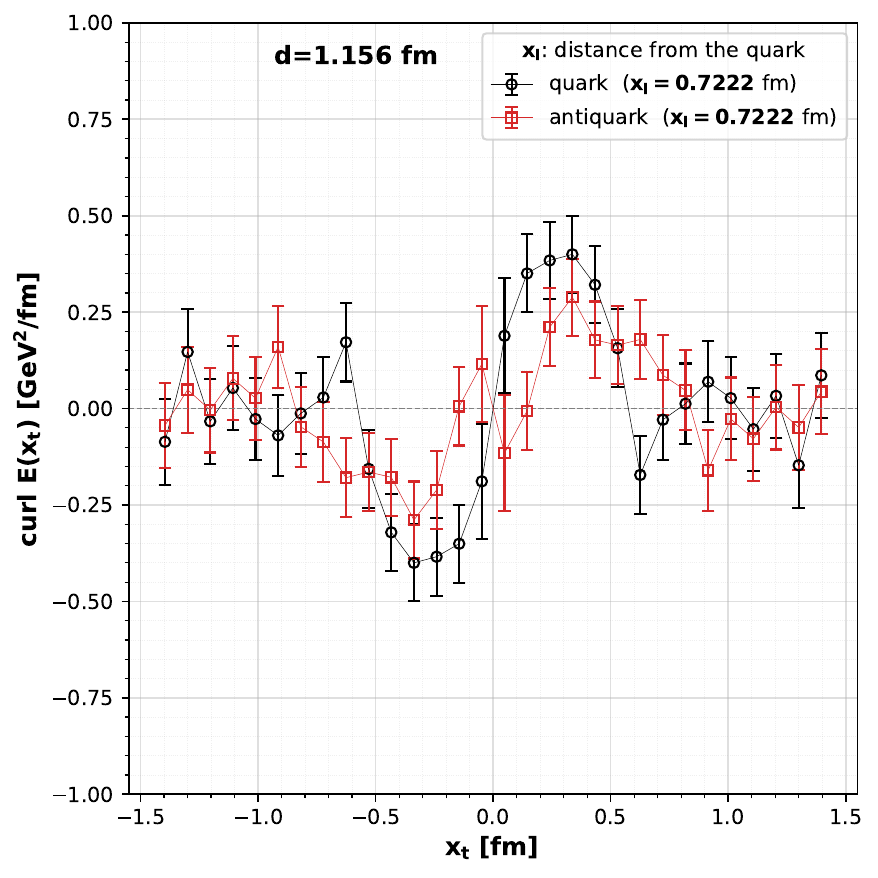}
\includegraphics[width=0.47\textwidth,clip]{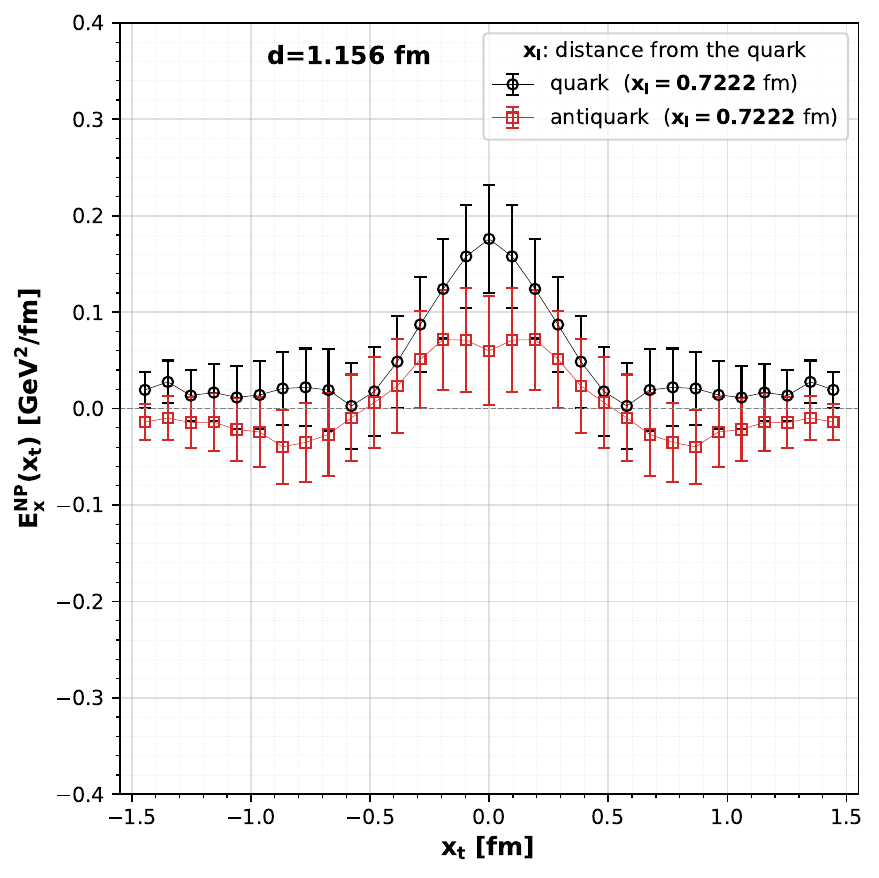}
%
\caption{\label{Fig3.7} Transverse distribution of the magnetic current and nonperturbative electric field
near the color sources and at the middle point for $d =12 \, a  \simeq  1.156 $ fm.}
\end{center}
\end{figure*}
\begin{figure*}
\begin{center}
\includegraphics[width=0.47\textwidth,clip]{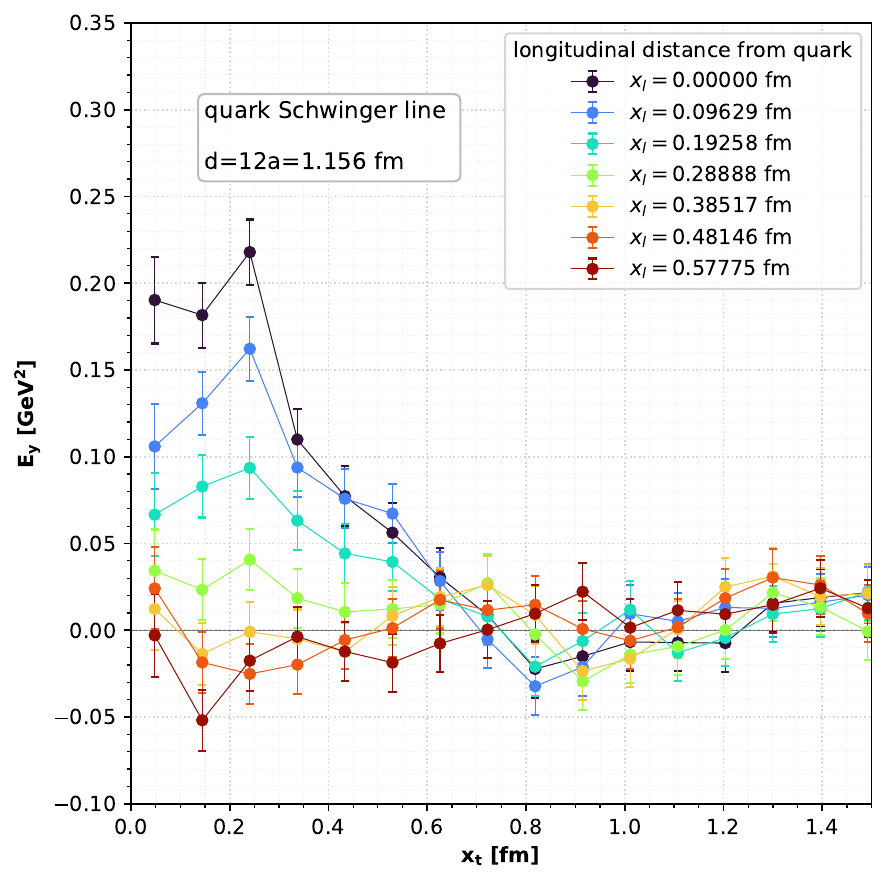}
\includegraphics[width=0.47\textwidth,clip]{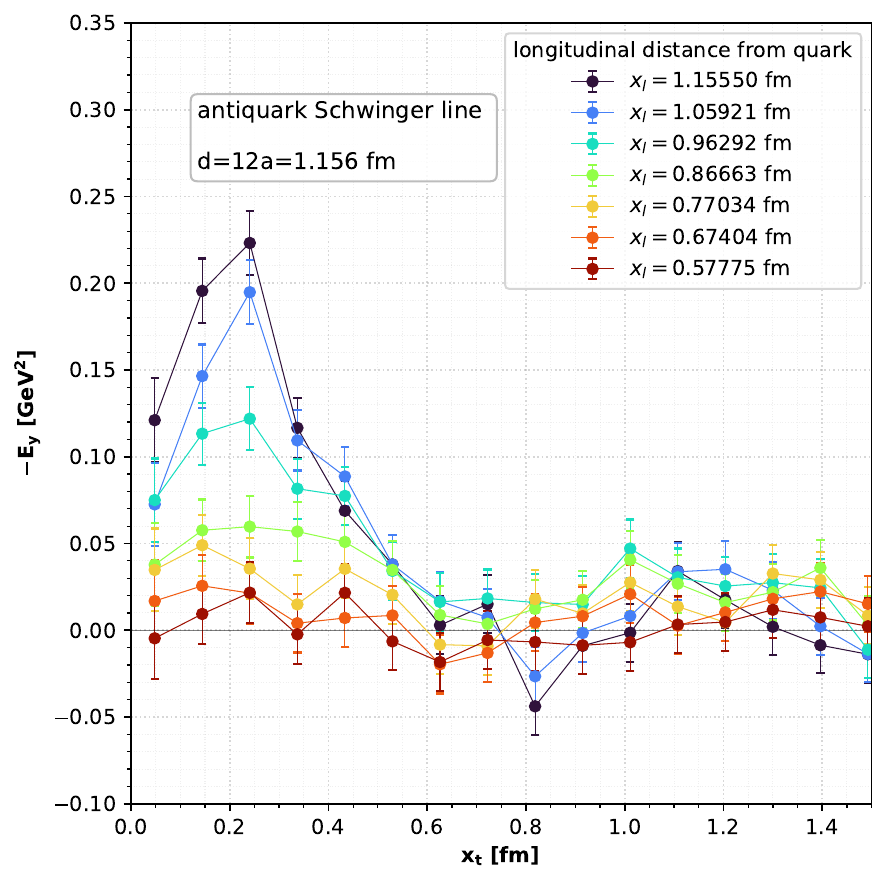}
\\
\includegraphics[width=0.47\textwidth,clip]{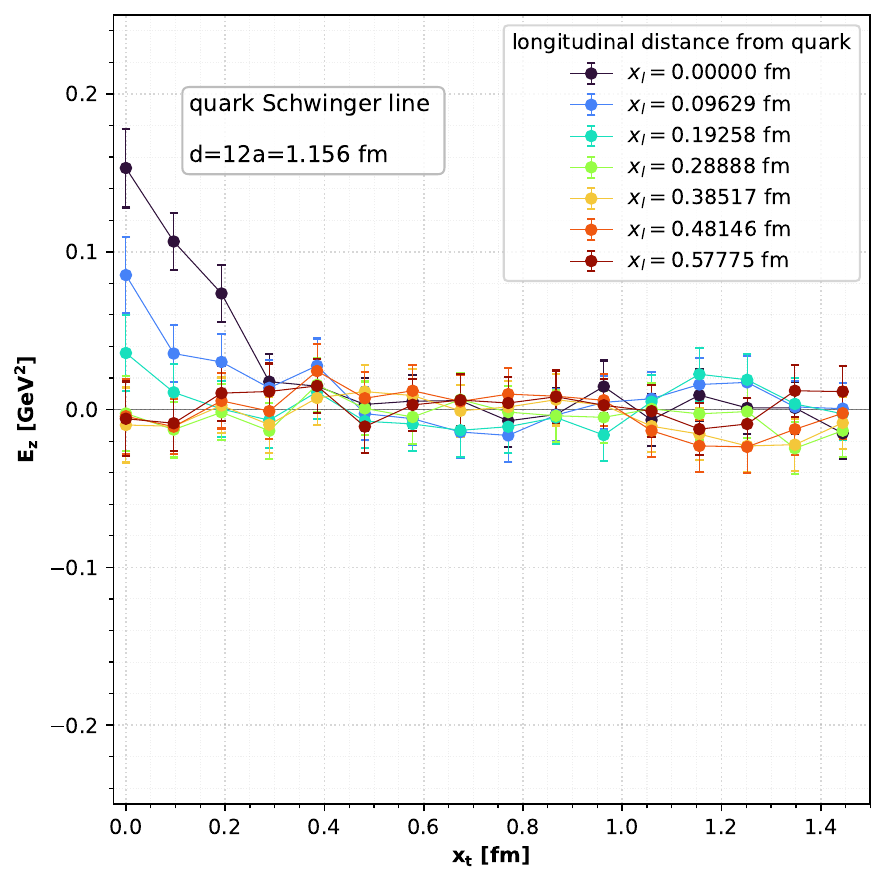}
\includegraphics[width=0.47\textwidth,clip]{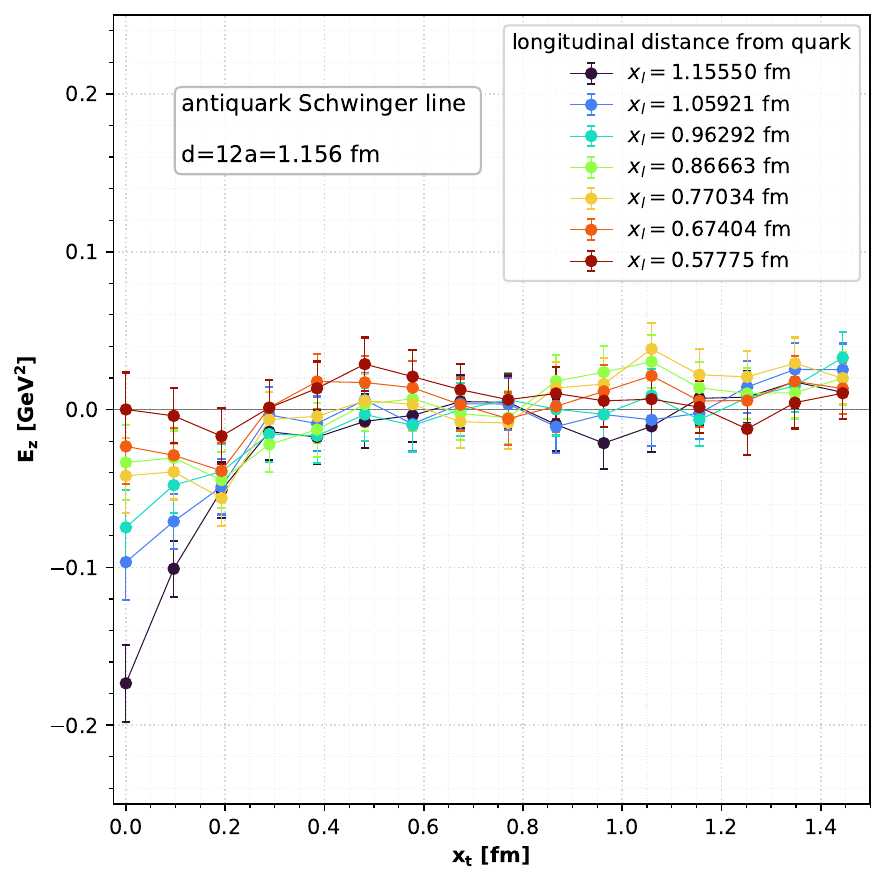}
%
\caption{\label{Fig3.8} Transverse distribution of the electric fields $E_y$ and $E_z$
along the flux tube for $d =12 \, a  \simeq  1.156 $ fm. }
\end{center}
\end{figure*}
\begin{figure}
\begin{center}
\includegraphics[width=0.47\textwidth,clip]{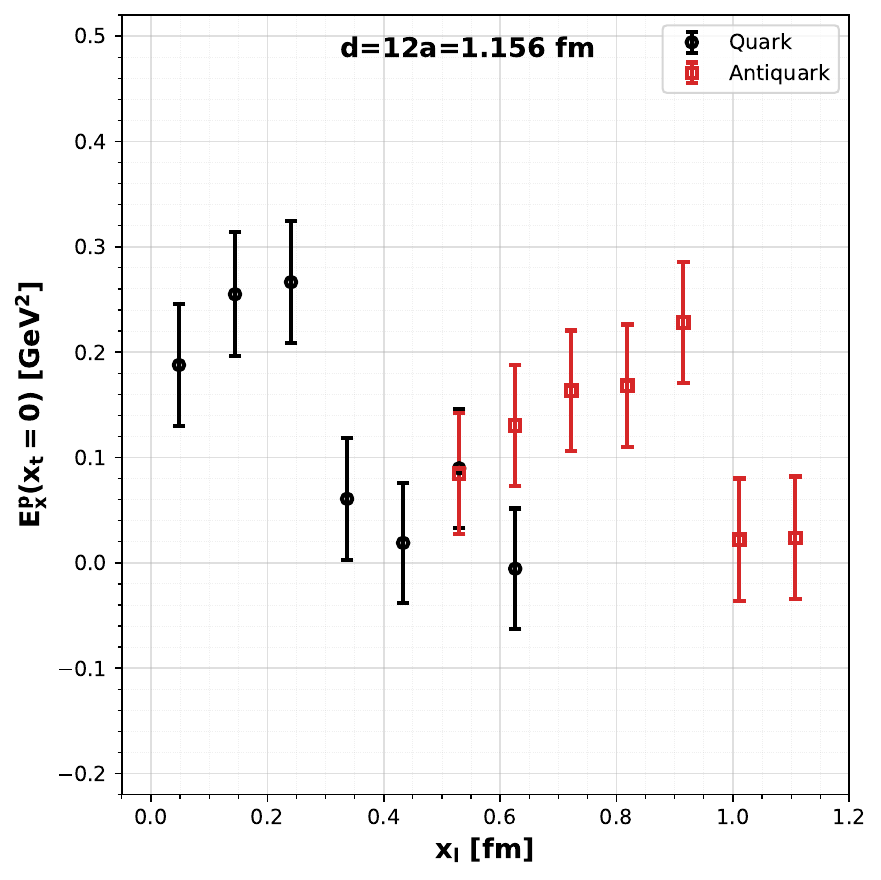}
\\
\includegraphics[width=0.47\textwidth,clip]{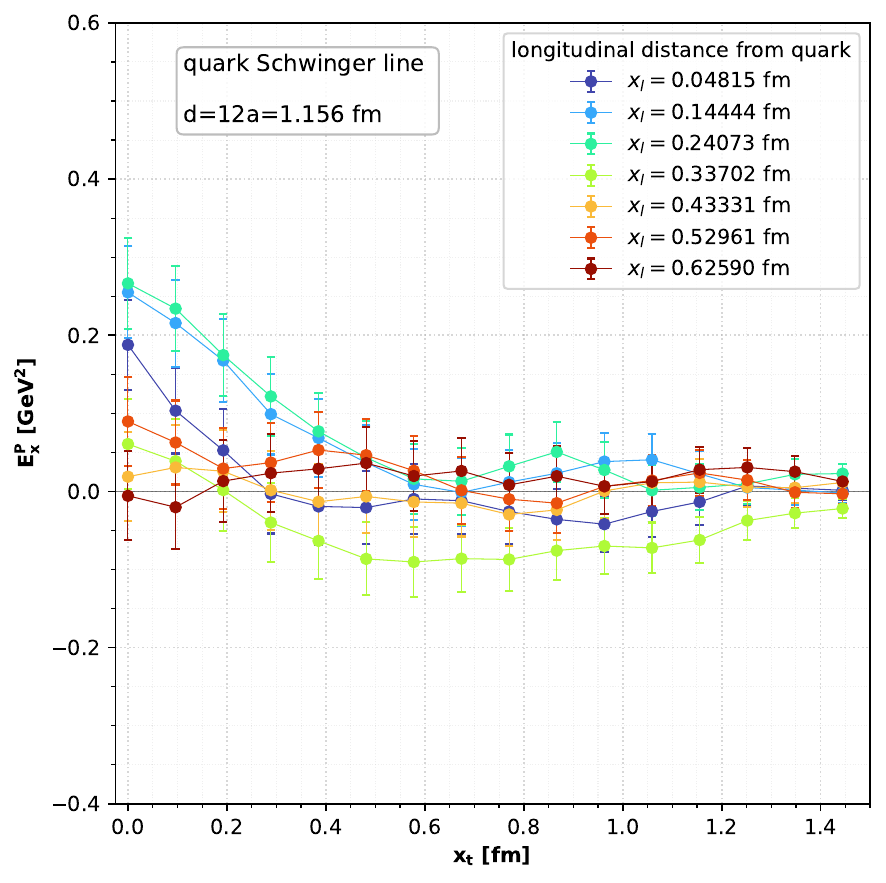}
\\
\includegraphics[width=0.47\textwidth,clip]{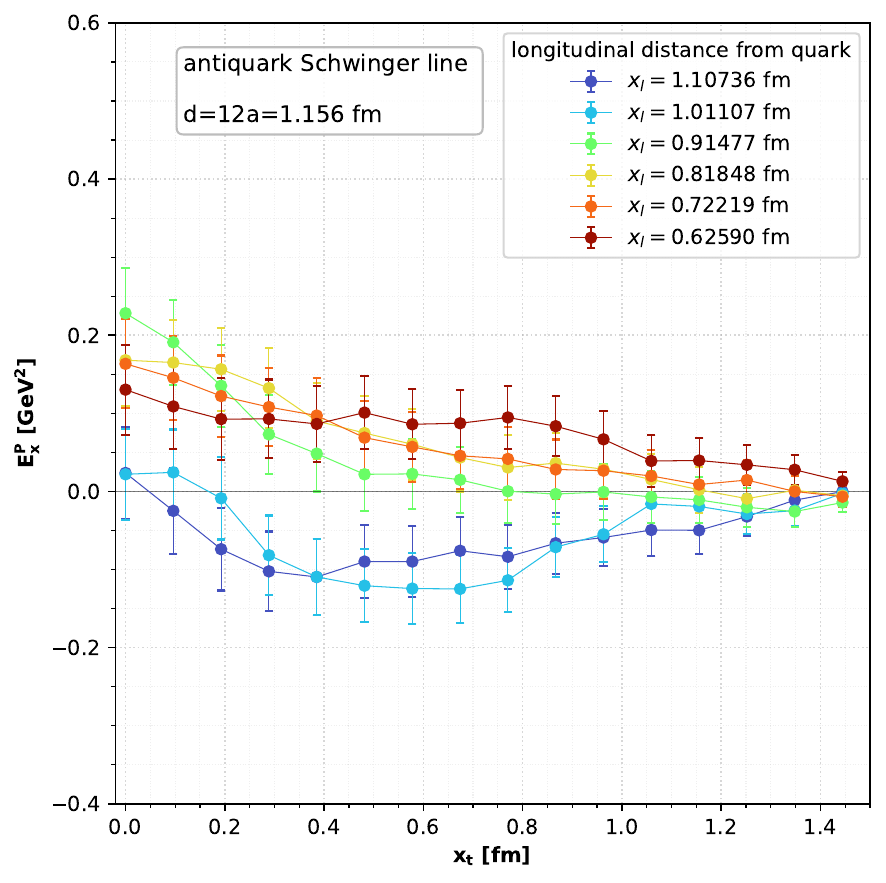}
%
%
\caption{\label{Fig3.9}  The perturbative longitudinal electric field $E^P_x$ at $x_t = 0$ {\it versus} the longitudinal distance $x_l$  (upper panel).
 Transverse distribution of   $E^P_x$  along the flux tube for $d =12 \, a  \simeq  1.156 $ fm (middle and lower panels).}
\end{center}
\end{figure}

 By increasing the distance $d$ between the static charges, it is expected that for $d > d^*$ the flux-tube string will break
  due to the dynamical generation of a quark-antiquark pair.  To detect the string breaking we may look at the
  breaking of the quark-antiquark  symmetry and/or antisymmetry and to the possible suppression  of the nonperturbative
  field around  the string-breaking point.  Accordingly, we considered the $12 \times 12$ Wilson loop at the same value of the lattice gauge coupling,
  $\beta = 6.880$, that corresponds  to static quark-antiquark color charges at distance $ d = 12 \, a \simeq 1.156 $ fm.
  Actually, in Fig.~\ref{Fig3.6}   the nonperturbative electric field is displayed at zero transverse distance $x_t = 0$ along the line joining the static color charges. 
 There is an indication  that the nonperturbative electric field profile does not satisfy anymore the quark-antiquark symmetry.
 In fact,  it seems that near the flux-tube middle point $E_x^{\rm NP}(x_t=0)$ obtained from the antiquark connected correlator is strongly suppressed
 with respect to the one from the quark connected correlator.  From this last  figure it seems that the flux tube breaks at around the middle point  leading to 
two strings with  $d_1 \approx   0.7$ fm and $d_2  \approx$  0.3 fm.  This is further confirmed by  Fig.~\ref{Fig3.7} showing
 the transverse distributions of the magnetic current and nonperturbative field near the color sources and at the middle, where there is almost no signal
 in the case of the antiquark connected correlator function and, moreover, the loss of the quark-antiquark symmetry manifests itself more convincingly. 
As a further check,  we looked at  the transverse fields $E_{y}$ and $E_{z}$.
In Fig.~\ref{Fig3.8}  we report the transverse distribution of the electric fields $E_{y}$ and $E_{z}$  in the first half (left panels) and second 
half (right panels) of the flux tube. Also here there are hints of the breaking of the quark-antiquark antisymmetry. More importantly, the transverse fields are no longer positive definite in the first half and negative definite in the second half of the line joining the sources, since in the region around the flux-tube middle point the transverse electric fields seem to reverse their sign.  This means that the transverse fields can no longer consistent with being the transverse component of the Coulomb field generated by a static 
quark-antiquark pair.  Since we have merely increased the source distance, the only  possible explanation is to admit the presence of additional color charges. 
Remarkably, a more stringent confirmation comes from  Fig.~\ref{Fig3.9}  where we present  
 the profile and transverse distributions of the  perturbative longitudinal field $E^P_x$.
 We see here a more evident breaking of the quark-antiquark symmetry and the fact that the longitudinal perturbative
field becomes negative near the string-breaking longitudinal distance.  Again, this forbids the interpretation of $E^P_x$  as the longitudinal component of
 the Coulomb field generated by only a static quark-antiquark pair.  \\
\subsection{Consistency checks}
\label{S3.3}
\begin{figure}
\begin{center}
\includegraphics[width=0.434\textwidth,clip]
{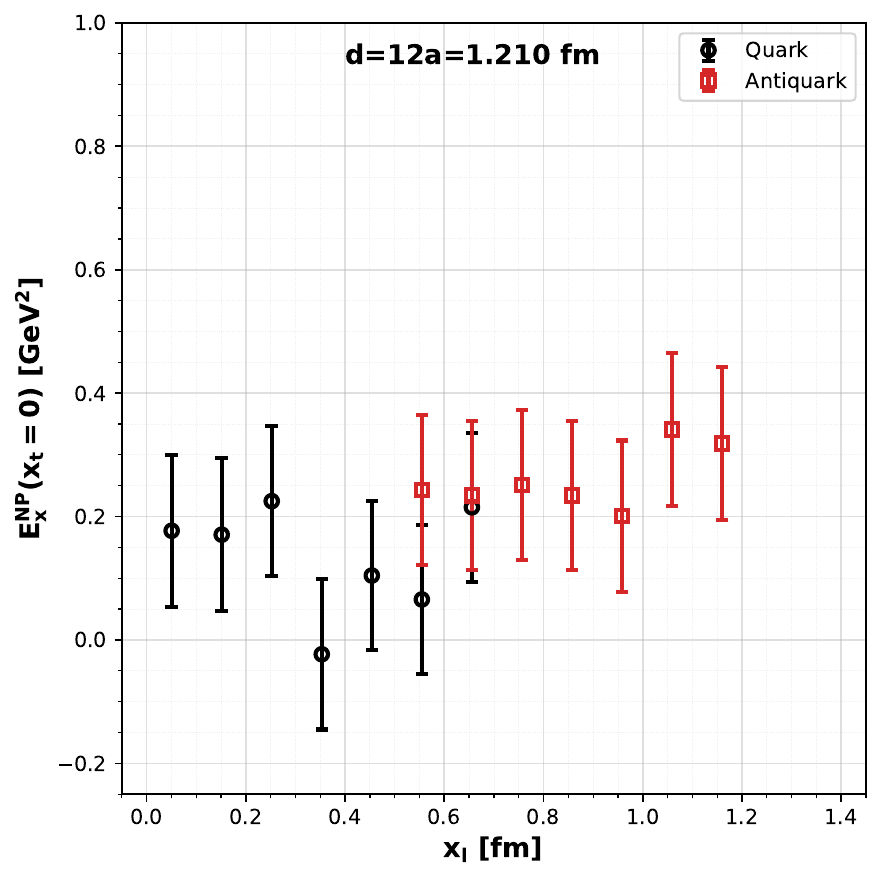}
\includegraphics[width=0.434\textwidth,clip]
{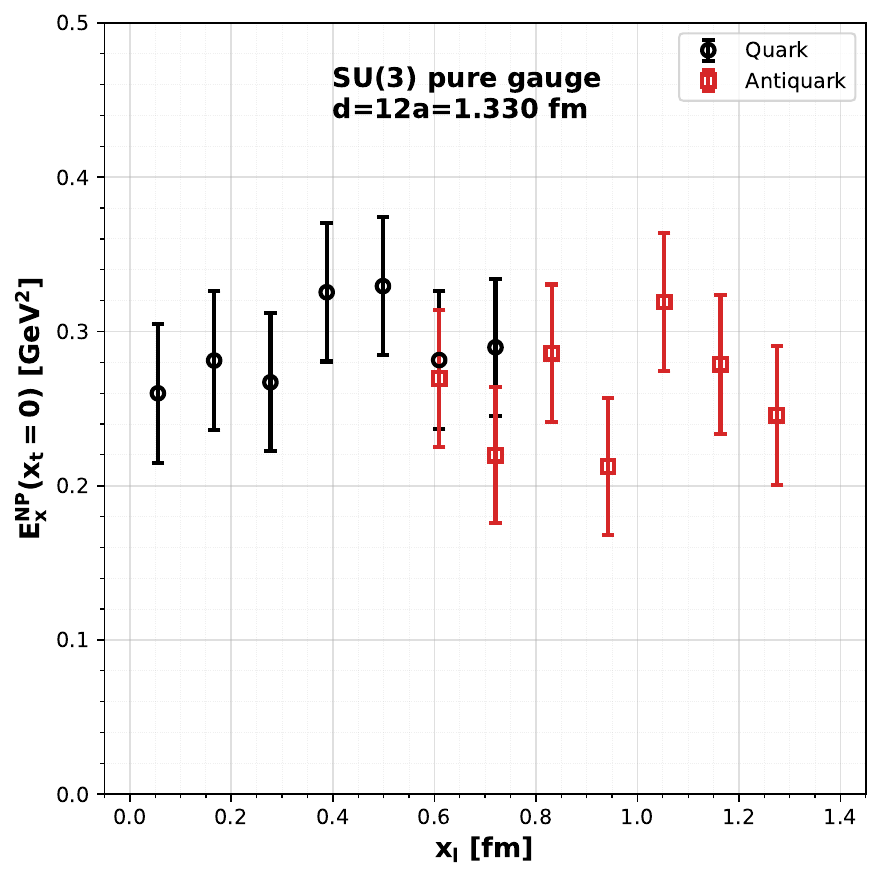}
%
\includegraphics[width=0.434\textwidth,clip]{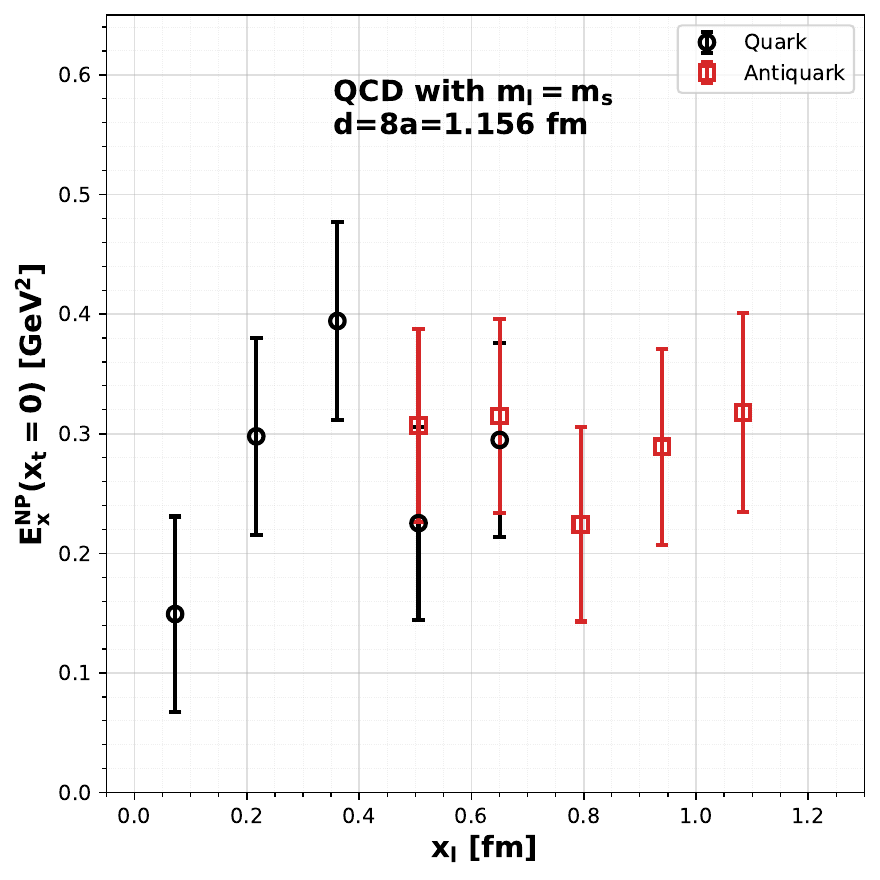}
%
%
\caption{\label{Fig3.10}  (upper panel) The nonperturbative longitudinal electric field $E_x^{\rm NP}$ at $x_t = 0$ {\it versus} the longitudinal distance $x_l$ for
QCD with (2+1) flavors of dynamical quarks at physical masses  $m_l \, = 1/27 \, m_s$  and static source distance  $d = 12 \, a  \simeq  1.210 $ fm. 
 (middle  panel)  $E_x^{\rm NP}(x_t = 0)$ {\it versus}  $x_l$  for the SU(3) pure-gauge  theory at  source distance $d = 12 \, a  \simeq  1.330 $ fm, and
 QCD with (2+1) flavors and symmetric quark masses $m_l \, = m_s$ with   $d = 12 \, a  \simeq  1.330 $ fm   (bottom panel). }
\end{center}
\end{figure}

In order to check the consistency of our hints of string breaking, we focus on the case of QCD at the physical point with $\beta = 6.832$ and 
distance  $d =12 \, a  \simeq  1.210$ fm  (see Table~\ref{measurements}). Since the distance between the static color sources is greater than previously considered,
we should still have string breaking. Indeed, looking at Fig.~\ref{Fig3.10}, upper panel,  where the profile of the nonperturbative electric field is plotted 
at zero transverse distance, we infer that the string seem to breaks at  about $x_l \approx 0.3$ fm into two smaller strings. This is also confirmed by
performing the same analyses as before of the transverse and longitudinal perturbative electric fields. These  conclusions hold also for
the larger distance   $d =14 \, a  \simeq  1.348$ fm,  $\beta = 6.880$ (see Table~\ref{measurements}) albeit with rather larger statistical uncertainness. \\
It is worthwhile to mention that our method can also shed light  on the physical mechanisms responsible for the string breaking in QCD. In fact,
it widely believed that the  QCD string breaking  can be developed in analogy with the production of $e^+  e^-$ pairs in a homogeneous electric field, 
the so-called Schwinger mechanism~\cite{PhysRev.93.615,PhysRev.82.664,PhysRevD.2.1191}, that  can be understood in this context as the $\bar{q} q$ 
 pair tunneling out of the vacuum.  
If  this is the case, then the tunneling amplitude should be suppressed by  increasing the quark masses. In particular,
 string breaking in quenched QCD, namely the SU(3) pure-gauge theory,  should be absent. Indeed, it is well known that in lattice pure-gauge theories
 there are ample evidences  of well-defined flux tube structures even for static quark-antiquark distances well beyond 1 fm. For instance, in our previous
  paper~\cite{Baker:2024peg}  in the case of  the SU(3) pure-gauge  theory we found that for a distance of  $d \simeq 1.235$ fm, {\it i.e.} well above $d \simeq 1.156$ fm, the signal of  the full longitudinal electric field    on the midplane was  clearly present. However, this result gives at best only an indication of the absence of string breaking
in the pure-gauge  theory. Therefore, we have repeated the  analysis by employing both the connected correlation functions. More precisely, we
have considered the SU(3) pure-gauge  theory at  three different distances of the static quark-antiquark color charges as summarized in 
Table~\ref{measurements}.
 Adopting the same methodology as in the previous subsection, we may safely confirm that for the pure-gauge  SU(3) theory  there is a 
well-behaved flux tube structure up to the greatest employed distance  $d \simeq 1.330$ fm.
 To show this,  we merely restrict ourself to display in Fig.~\ref{Fig3.10}, middle  panel, the nonperturbative longitudinal electric field
 $E_x^{\rm NP}$ at $x_t = 0$ along the flux tube  for source distance $d = 12 \, a  \simeq  1.330 $ fm. \\
Interestingly, our approach allows to detect the possible dependence of the
 string-breaking distance $d^*$ on the quark masses in QCD with dynamical quark fields. To do this, we have also considered 
 QCD  with (2+1) flavors of dynamical quarks with symmetric masses,  $m_l = m_s$.  To increase the signal strength over the statical noise, we  used   
 $ \beta = 6.472$,  corresponding to  $a(\beta)  \simeq  0.14452 $ fm ,  and  $d = 8 \, a  \simeq  1.156 $ fm  (see Table~\ref{measurements}).
In Fig.~\ref{Fig3.10}, bottom panel, we show the nonperturbative longitudinal electric field  $E_x^{\rm NP}$ at $x_t = 0$ as a function of the longitudinal coordinate $x_l$ for
QCD with symmetrical quark masses. From this figure it is quite evident that there is an almost uniform nonperturbative longitudinal electric field
along the whole region between the static color sources that, in addition, respects the quark-antiquark symmetry. Obviously, this last statement is also
confirmed by looking at the transverse and longitudinal perturbative electric fields and the transverse distributions of the magnetic current and nonperturbative 
electric field along the flux tube.   As a consequence, we can conclude that for  $d  \simeq  1.156 $ fm  there is no string breaking at 
$m_l = m_s$, while for physical quark masses ($m_{\pi} \simeq 140$ MeV) our previous analyses support evidences for string breaking. 
This gives a hint that the string-breaking distance $d^*$ increases with the quark masses. 
\section{Summary and Conclusions}
\label{S4}
In this paper we have studied the flux-tube electric fields for various different distances  between static quark-antiquark color charges.
This investigation has been performed mainly  in (2+1)-flavor QCD, but we have also considered the SU(3) pure-gauge  theory.
Our main aim was  to investigate string breaking  in QCD by means of a direct and model-independent approach. 
The natural path to detect the string breaking on the lattice is to use the Wilson loop 
 as an observable. Indeed, the breaking of the string should manifest itself
 by the potential becoming constant above a certain threshold distance. 
 However, we already pointed out that  this phenomenon could not be clearly observed 
 so far  due to the fact that
 the Wilson loop operator has a very small overlap with the ground state after 
 the string is broken, thus suggesting that the string breaking is a mixing phenomenon. 
 As a matter of fact, a model that gives an explicit picture of string breaking in the presence of dynamical 
 quarks as a mixing phenomenon that involves the string state and a two-meson state has been presented
 long ago. This leads to the consequence  that  to detect the string breaking and  estimate the string-breaking distance in QCD with
 dynamical quarks one must inevitably make some theoretical assumptions. \\
 The approach discussed in the present paper  allows to obtain  a reliable model-independent criterion to enlighten  string breaking in full QCD
by looking directly at the  gauge-in\-va\-riant longitudinal electric field in the region between  two static sources.  
The main advantage of this method resides on the fact that, by employing  the two connected correlators with   
the Schwinger line  is attached to the quark time line or to the antiquark time line,  we can look directly at both the  nonperturbative gauge-invariant
longitudinal electric field $E_x^{\rm NP}$ and the perturbative Coulomb-like electric fields  $E_{y}$, $E_{z}$ and $E^P_x$,  in the whole region between the two static color sources responsible for the formation of the flux-tube structure.
Our approach has led us to  directly  observe the phenomenon of the string breaking in QCD with dynamical quark fields. In particular,
 we were able to constrain the so called string-breaking distance $d^*$ within a rather small interval (0.963 fm to 1.116 fm) for  QCD with (2+1) flavors
 of  dynamical quarks with physical masses. 
 We have also checked that  string breaking is consistent with the usually assumed Schwinger mechanism since the string-breaking distance
 increases with the quark masses. 
\section*{Acknowledgments}
We wish to express our deep gratitude to the late Marshall Baker, whose scientific insight, discussions, and contributions to the development of the work presented in this paper were invaluable.

This investigation was in part based on the MILC collaboration's public lattice gauge theory code (\url{https://github.com/milc-qcd/}). Numerical calculations have been made possible through a CINECA-INFN agreement, providing access to HPC resources at CINECA. PC, LC and AP acknowledge support from INFN/NPQCD project. VC acknowledges support by  the Deutsche Forschungsgemeinschaft \linebreak (DFG, German Research Foundation) through the CRC-TR 211  ``Strong-interaction matter under extreme conditions''  -- \linebreak project number 315477589 -- TRR 211. This work is (partially) supported by ICSC - Centro Nazionale di Ricerca in High Performance Computing, Big Data and Quantum Computing, funded by European Union - NextGenerationEU.

\bibliography{qcd}

\end{document}